# JOINT SURVEY PROCESSING STUDY: MAR 2019 FINAL REPORT

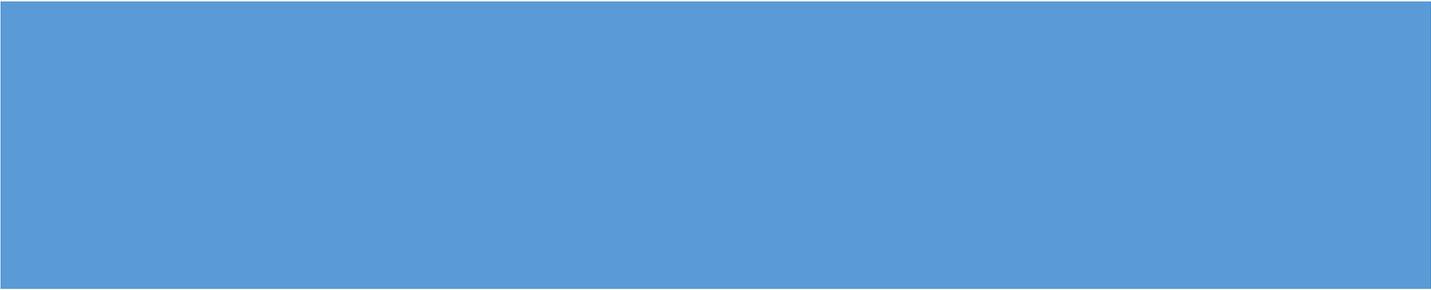

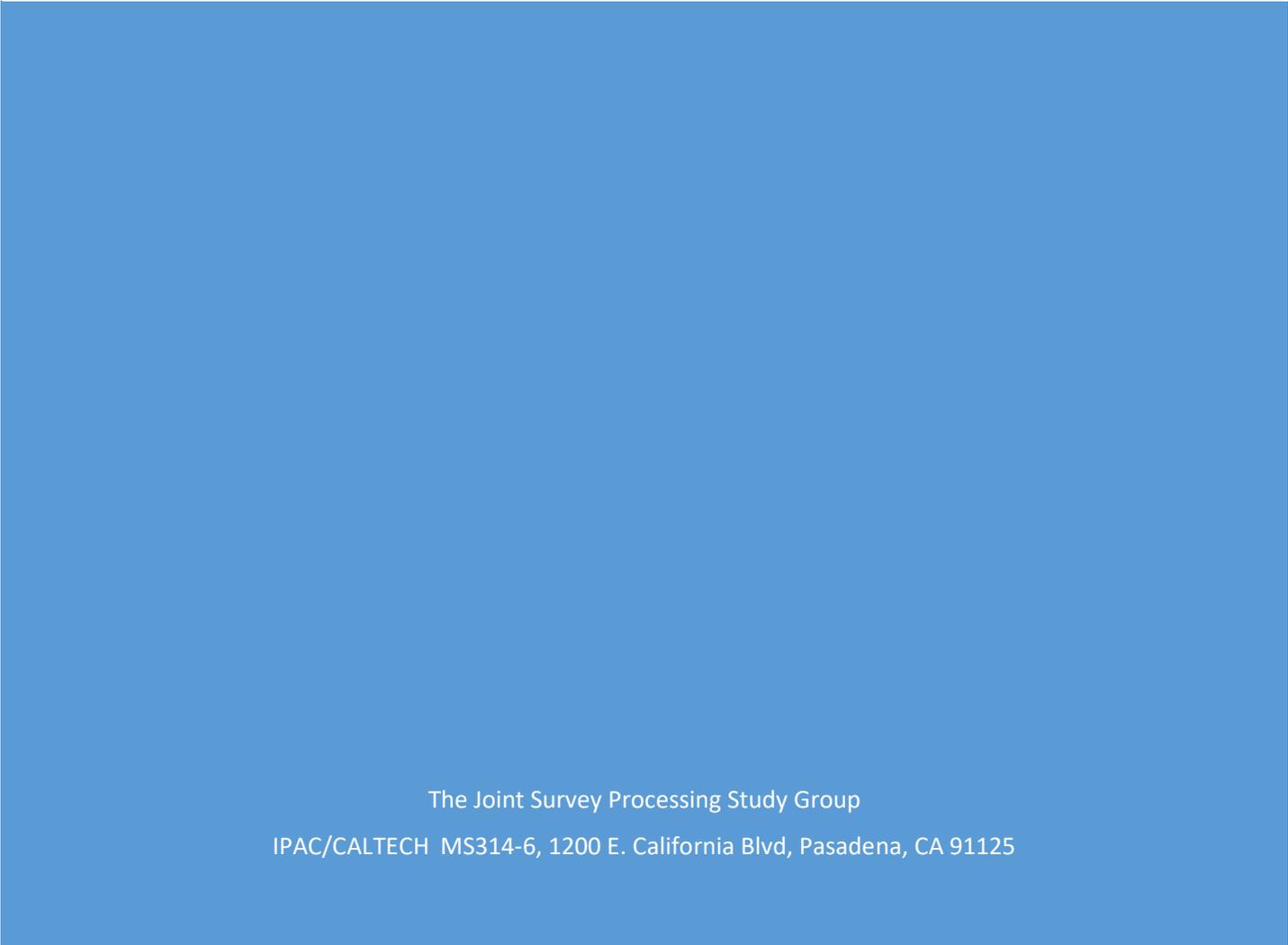

The Joint Survey Processing Study Group

IPAC/CALTECH  MS314-6, 1200 E. California Blvd, Pasadena, CA 91125

# The Joint Survey Processing Study Group

**Coordinators:**
Ranga Ram Chary rchary@caltech.edu & George Helou gxh@ipac.caltech.edu

**Co-authors:**
Gabriel Brammer: gbrammer@gmail.com
Peter Capak: capak@ipac.caltech.edu
Andreas Faisst: afaisst@ipac.caltech.edu
Dave Flynn: dflynn@ipac.caltech.edu
Steven Groom: sgroom@ipac.caltech.edu
Henry C. Ferguson: ferguson@stsci.edu
Carl Grillmair: carl@ipac.caltech.edu
Shoubaneh Hemmati: shemmati@caltech.edu
Anton Koekemoer: koekemoer@stsci.edu
BoMee Lee: bomee@ipac.caltech.edu
Sangeeta Malhotra: sangeeta.malhotra@nasa.gov
Hironao Miyatake: hironao.miyatake@iar.nagoya-u.ac.jp
Peter Melchior: melchior@astro.princeton.edu
Ivelina Momcheva: imomcheva@stsci.edu
Jeffrey Newman: janewman@pitt.edu
Joseph Masiero : Joseph.Masiero@jpl.nasa.gov
Roberta Paladini: paladini@ipac.caltech.edu
Abhishek Prakash: aprakash@ipac.caltech.edu
Benjamin Rusholme: rusholme@caltech.edu
Nathaniel Stickley: nrstickley@ipac.caltech.edu
Arfon Smith: arfon@stsci.edu
Michael Wood-Vasey: wmwv@pitt.edu
Harry I. Teplitz: hit@ipac.caltech.edu

**Observers:**
Bruno Altieri: bruno.altieri@sciops.esa.int, Phil Appleton: apple@ipac.caltech.edu, Lee Armus: lee@ipac.caltech.edu, Bruce Berriman: gbb@ipac.caltech.edu, Sean Carey: carey@ipac.caltech.edu, Ken Carpenter: Kenneth.G.Carpenter@nasa.gov, Roc M. Cutri: roc@ipac.caltech.edu, William Anthony Dawson: dawson29@llnl.gov, Michael Fox: mfox@stsci.edu, Steve Kahn: skahn@lsst.org, Davy Kirkpatrick: davy@ipac.caltech.edu, Jeffrey Kruk: jeffrey.w.kruk@nasa.gov, Kian-Tat Lim: ktl@slac.stanford.edu, Robert Lupton: rhl@astro.princeton.edu, Julie McEnery: julie.e.mcenery@nasa.gov, Jason Rhodes: jason.d.rhodes@jpl.nasa.gov, Marc Sauvage: marc.sauvage@cea.fr, Michael D. Schneider: schneider42@llnl.gov, Roeland van der Marel: marel@stsci.edu, Stefanie Wachter: wachter@ipac.caltech.edu

# Table of Contents

## Executive Summary

*"The whole is greater than the sum of its parts"*

The Euclid, LSST and WFIRST projects will undertake flagship optical/near-infrared surveys in the next decade. By mapping thousands of square degrees of sky and covering the electromagnetic spectrum between 0.3 and 2 microns with sub-arcsec resolution, these projects will detect several tens of billions of sources, enable a wide range of astrophysical investigations by the astronomical community and provide unprecedented constraints on the nature of dark energy and dark matter. The ultimate cosmological, astrophysical and time-domain science yield from these missions will require "joint survey processing" (JSP) functionality at the pixel level that is outside the scope of the individual survey projects. The JSP effort scoped here serves two high-level objectives: 1) provide precise concordance multi-wavelength images and catalogs over the entire sky area where these surveys overlap, and 2) provide a science platform to analyze concordance images and catalogs to enable a wide range of astrophysical science goals to be formulated and addressed by the research community.

In support of these objectives, we have scoped the development of the following capabilities:
1)    Reconcile and standardize the astrometry and photometry of the image-level data products and associated metadata available from each of the three surveys;
2)    Leverage these standardized products to analyze the survey data jointly at the pixel level to generate the ultimately precise, de-confused, extinction-corrected photometric catalogs. Through fake source simulations, these catalogs will be well characterized in photometric uncertainty, completeness and reliability over the overlapping sky areas.
3)    Build and maintain a science platform, including data access and manipulation tools, to be deployed in various computing environments to enable and support:
   a.    Generation of similar catalogs combining with other ancillary data sets that will be obtained in support of Euclid and WFIRST, e.g. with CFHT, PanSTARRS, Hyper-SuprimeCAM,
   b.    Astronomical analysis (e.g. shape fitting, diffuse emission extraction, multi-wavelength model fitting) of the standardized, calibrated frames from Euclid/LSST/WFIRST,
   c.    Integration of the standardized Euclid/LSST/WFIRST catalogs to ancillary, lower-resolution, all-sky data products such as GALEX and WISE;
4)    Offer a networking environment facilitating access to high performance computing that will enable algorithmic manipulation and analyses of these data and Monte Carlo simulations to be performed by the community in support of these analyses.

It is most cost effective to develop these capabilities as a coherent, robust service for all users by building on the insights and expertise of the three survey projects working closely with a dedicated JSP Core Team. With these capabilities available, JSP will allow the U.S. (and international) astronomical community to manipulate the flagship data sets and undertake innovative science investigations ranging from solar system object characterization, exoplanet detections, nearby galaxy rotation rates and dark matter properties, to epoch of reionization studies. It will also allow for the ultimate constraints on cosmological parameters and the nature of dark energy, with far smaller uncertainties and a better handle on systematics than by any one survey alone.

In this report we examine the science case for Joint Survey Processing, sketch the data system and inventory the software needed to address the science case, and derive preliminary technical requirements. We then describe an architecture capable of meeting those requirements, along with the software tools and services, release schedule and computing needs. We list the tasks required to build the system and deliver the products and tools, and scope the effort required for each from 2019 to 2032. We estimate a total of 204 WY, 69% of which is allocated to the Joint Survey Processing Core Task, 15% to the three individual survey projects, and 16% to the community, funded through competitive proposal cycles. The 204 WY number is largely driven by the duration of the effort. During the five peak years, the effort averages about 24 WY/year for all teams and sources of funding. In addition to the workforce, we estimate a billion or more CPU hours will be needed. A reasonable scenario would be for DOE to allocate that computing load on their Supercomputing Centers, or procure the equivalent on the commercial cloud if that became competitively priced in the next few years and comparable network speeds can be demonstrated.

# 1. Timelines and Technical Motivation: The Time is Now!

Euclid is scheduled to launch in June 2022 while LSST is planning to start science operations on a similar time frame (Table 1). Deep (>25$_{th}$ mag), multi-band, optical photometry from LSST and other current ground-based telescopes is crucial to estimate precise photometric redshifts of galaxies whose shear distortion is being measured by Euclid and LSST. These are essential for measuring the three-dimensional weak-lensing power spectrum, which leads to constraints on the temporal evolution of dark energy, one of the primary goals of Euclid. LSST in turn relies on the deep near-infrared data from Euclid for improved photometric redshifts and for the ~0.15" spatial resolution optical data (in a broad visible band spanning 550-900nm) for joint shear analysis.

The Euclid wide area survey has no repeats while LSST will cover a significant fraction of the observable sky every few nights (Figure 1). Thus, right from the beginning, there is expected to be complete coverage of the Euclid-observed sky (5sigma VIS~25.2 AB mag) by LSST although the single-epoch LSST data will be relatively shallow in the initial stages with 5sigma~24.5 AB mag, but achieving comparable depth within a year (Table 1). Furthermore, the deep field data (~40 deg$_2$) from both LSST (full depth 28.6 AB mag) and Euclid (full depth 26.5 AB mag) will be co-spatial from the earliest stages, with the sensitivity continuously increasing with time. However, the Galactic Plane and Ecliptic Plane will remain unobserved by Euclid until the end of the prime survey which will be in 2028. WFIRST on the other hand is expected to undertake dedicated surveys of the inner Galaxy starting with the first year after its launch for the microlensing survey.

Despite the avoidance by LSST of the Galactic bulge fields in the early years, at these depths, source confusion is highly relevant when the seeing resolution is ~0.8-0.9" FWHM (Figures 2, 3 and 4) with up to 25% of sources affected in the first year stacks of LSST and up to 50% affected in the full-depth stacks. This implies that even in the early stages of both the LSST and Euclid projects (~2022-2023), joint survey processing (JSP) at the pixel level will be crucial for optimal photometry/spectrophotometry of sources and galaxy shape measurements. This will entail: 1) generating astrometrically aligned multi-wavelength images with alignment across projects matching that within a project (typically ~10mas/epoch); and 2) using knowledge of brightness, morphology and position from high spatial resolution (<0.3") space-based data, in conjunction with an accurate estimation of the PSF, for source extraction in the ground-based data.

**Development of the infrastructure for joint survey processing (JSP) is timely and urgent. The first Euclid data release is expected in mid-2023, 14 months after launch while the first LSST release is scheduled slightly sooner, 12 months after the start of operations. With such a short timeline for data release, each of the projects will be overwhelmed with handling the systematics from their individual datasets prior to release. The Joint Survey Processing Team (authors listed on the cover page) has expertise straddling LSST, Euclid and WFIRST; thus, the team can build in parallel the capabilities required to ingest the data products delivered by the projects and through joint processing, provide a value-added science-quality product that will enable more accurate astrophysical and cosmological investigations, which the individual datasets alone will not be able to support.**

In addition, the access to these data sets through a unified interface (or science platform) would be equivalent to a virtual powerful optical/infrared observational facility and allow community users to interactively manipulate the datasets to achieve their own science goals. In order to improve the fidelity of scope estimates, this team has obtained sample data of quality comparable from Euclid, LSST and WFIRST through existing HST and Subaru surveys. These are in the 2 square degree COSMOS field, and the prototyping has laid the foundations for full development of joint survey processing capabilities on a timeline that will be ready for the first data releases.

**Table 1: Comparison between the different large optical/NIR surveys of the 2020s**

| Project | Area deg² | 5σ Sensitivity AB mag | Spatial resolution (FWHM) | Pixel Scale | Year |
|---|---|---|---|---|---|
| *Euclid VIS* | 15000 | VIS~25.2 | 0.15" | 0.1" | 2022-2028 |
| *Euclid NIR* | 15000 | Y,J,H~24 | 0.3-0.6" | 0.3" | 2022-2028 |
| WFIRST-HLS | 2200 | Z,y,J,H, F184~26.7 | 0.18" | 0.11" | 2026- |
| LSST | 20000 | u~26.1,g~27.4,r~27.5,i~26.8,z~26.1,y~24.9 | 0.81" | 0.2" | 2022-2032 |

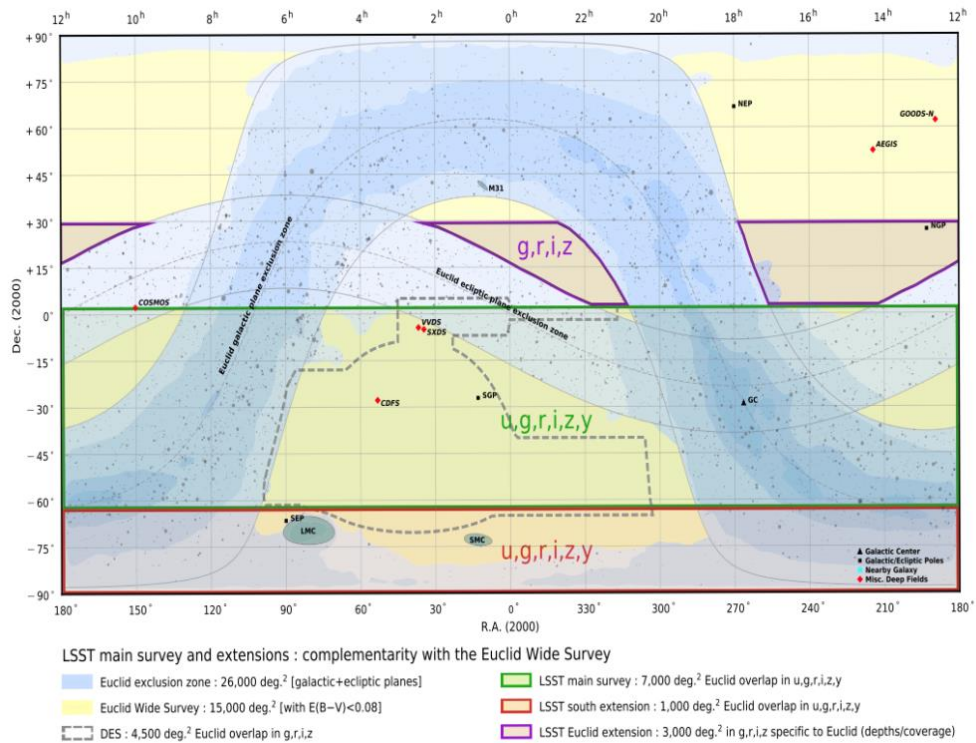

Figure 1: The sky coverage of the Euclid wide-area survey (yellow) and the possible LSST northern extension (purple rectangle from Rhodes et al. 2017). LSST, will in addition, cover the entire 20000 deg₂ Southern hemisphere to ~27 AB mag over 10 years (green and red rectangles).

The WFIRST high latitude survey field (2200 deg₂) is a part of the yellow region between RA of 300 and 90 degrees in the Southern sky while the WFIRST microlensing survey will be in the vicinity of the Galactic Center (labeled GC at 270,-30). Figure courtesy of Jean-Charles Cuillandre/Euclid project. The large overlap in sky coverage between these three projects and the desire for matched precision photometry, motivates a joint analysis of the data.

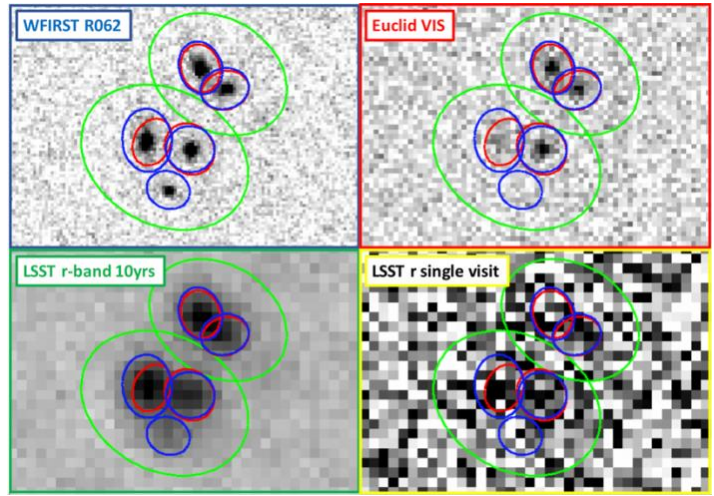

Figure 2: An illustration of source confusion in an optical band (centered at ~6000A) from the three primary surveys, along with the isophotes derived from photometry on each of the images. The green isophotes are derived from the LSST r-band full-depth data of 27 AB mag, the red isophotes are from the Euclid only VIS data while the blue isophotes are for the deeper WFIRST data (Table 1). The sources are barely detected in the LSST single epoch data. In the absence of the deeper, space-resolution data, source confusion would result in both erroneous shape and photometry estimates in LSST data and also affect catalog matching. Conversely, both Euclid and WFIRST rely on deconfused optical photometry from LSST to get reliable photometric redshifts for galaxies that are detected in their respective surveys.

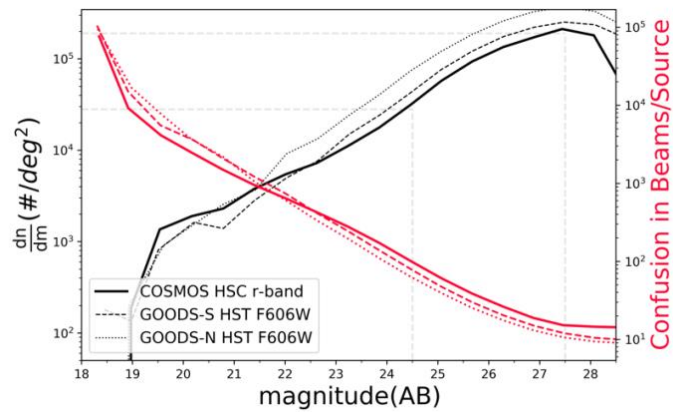

Figure 3: Visible band extragalactic source counts in the COSMOS, GOODS-S and GOODS-N fields shown as the black lines with the corresponding beams/source, a parameter of confusion, show as the red lines and corresponding to the right axis. Beams/source=40 (corresponding to ~25 AB mag) is the classical confusion limit. Using prior information on position and morphologies of sources allows one to reliably extract sources down to ~8 beams/source depending on the accuracy of the PSF and the quality of the priors (Magnelli et al. 2009, 2011). Since LSST has a single epoch

photometric sensitivity of 24.5 AB mag, with ~200 visits per band over 10 years, the images will start to be affected by confusion, within a year of LSST operations, if the spatial resolution has a median seeing of 0.8-0.9" FWHM. This implies that between 10-25% of galaxies will have their photometry affected by confusion in 1 year stacks of LSST data with up to 50% affected in the full depth stacks.

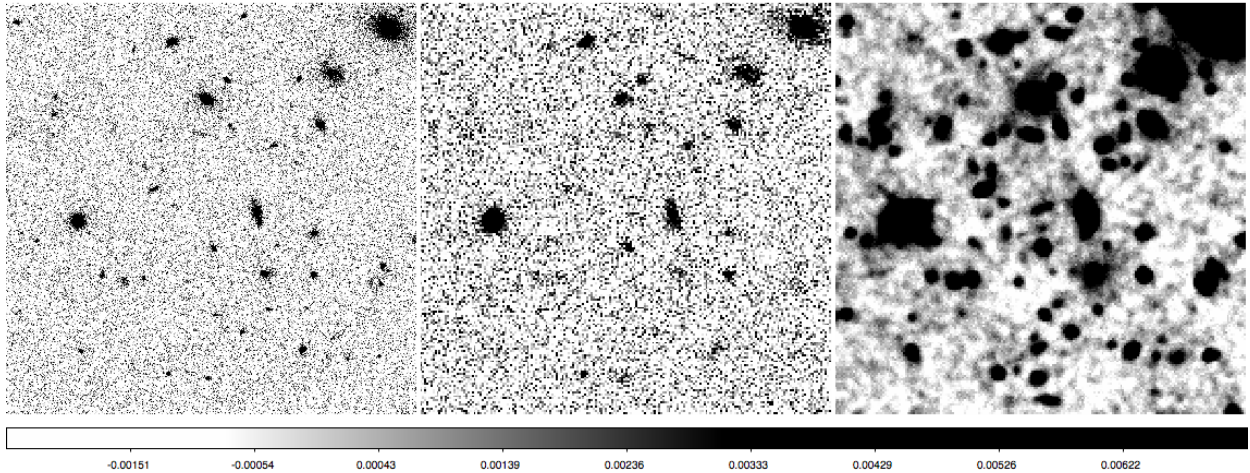

Figure 4: Simulated 40"x40" WFIRST (R602), Euclid (VIS) and LSST (full-depth r-band) images illustrating the variation in source density and source area across these surveys, which precludes high precision photometry through catalog matching. A pre-requisite for deriving precise photometric redshifts is photometry using tools like TRACTOR/TPHOT/SCARLET (Lang et al., Merlin et al. 2015), which jointly fit low- and high-resolution images for optimal photometric and morphology measurements. Joint survey processing will undertake this task, taking the final data products from the individual survey projects, standardizing them and developing a value-added optical/NIR dataset which will enable a range of science investigations beyond the capability of any single project.

## 2. Scientific Motivation

The combined analyses of optical and near-infrared datasets taken at spatial resolutions of 0.1"-0.8", over a time frame of >10 years, benefit *all* branches of astrophysics. A combination of data products from different imaging surveys is traditionally performed at the level of source catalogs. This approach is beneficial as long as the catalogs can unambiguously be matched and the measured photometry across varying PSF sizes, seeing conditions and astrometry calibrations can be accurately and unambiguously combined. However, at the depth and spatial resolution of LSST (or any large-area ground-based survey), where variable (and occasionally poor) seeing is a significant factor, the mapping is far less certain, increasing the uncertainty on information extracted from multiple surveys. The resulting uncertainty adds to the systematics floor of high-precision science experiments. We argue that the joint analysis at the pixel-level of the optical and near-infrared datasets from LSST, Euclid, and WFIRST, taken at spatial resolutions of 0.2"-0.8" over a time frame of > 10 years, will remedy these shortcomings.

It is crucial to note that none of these tasks proposed here is intended to conflict with the tasks that the individual survey projects have already been funded to undertake. If such tasks are already being undertaken by the respective consortia, the goal here is to build the infrastructure and common standards for combining and jointly analyzing the data. In what follows, we do not assume privileged data access; all operations described here can be performed on planned and publicly accessible data products.

## 2.1 COSMOLOGY

The measurement of the energy density of dark energy and its evolution with cosmic time is one of the key goals of the three projects. The reduction in uncertainty in the estimated redshift evolution of the dark energy equation of state is one of the topics that benefits dramatically from a form of joint analysis, with a factor of 2 to 3 improvement on the uncertainty in w' (the temporally varying dark energy equation of state) and a better handle on systematics by combining the cosmological parameters derived from different techniques (Jain et al. 2015, Rhodes et al. 2017). The reduction in uncertainty arises from: 1) improved photometry which translates to better photometric redshifts and reduced outlier (catastrophic redshift failures) rates, 2) joint shear analyses between LSST and Euclid or WFIRST for weak-lensing studies, 3) breaking redshift degeneracies in grism spectra for BAO studies and 4) the selection of high quality Type Ia SNe based on the properties of their host galaxies. Furthermore, independent constraints on the Hubble parameter can be derived from measurement of time-delays in strongly lensed, time-varying sources.

1.    Improving the precision of galaxy redshifts for 3D weak lensing and BAO studies by improving galaxy multi-wavelength photometry (see Figure 5, 6);

This could be done by using priors on morphologies and positions from the space-based datasets to extract photometry in the ground-based data. In addition, if a high-accuracy standardization of the data sets can be achieved, one can simultaneously model all available images, from all surveys, for each respective source (Figure 7). It is worth noting that the morphological k-correction (i.e. difference in morphology because of sampling different rest-frame wavelengths) between the EUCLID VIS band (which spans 550-900nm) and LSST should be small; however, there could be significant differences in morphology between the space NIR bands and LSST due to the relative importance of bulge and disk-like structures in the galaxies which multi-component galaxy models can account for. The main advantage of pixel-level analysis therefore, is the cross-band consistency of photometry and alleviation of photometric and shape contamination by source confusion in the deep, seeing-limited LSST dataset, especially when approaching its full 10-year depth of i~27.

a.    Breaking redshift degeneracies with improved photometry will also help with line misidentification in the spectral data. This is particularly important since at the spectral resolution of the Euclid grism and the typical line SNR of ~3.5, H$\alpha$+[NII] will be blended and could be misidentified as other lines, such as the [OIII] doublet or H$\beta$. Since the equivalent width of nebular emission lines is increasing with increasing redshift, a detected line could potentially be from a higher redshift source, i.e. a 0.9<z<1.8 H$\alpha$ emitter could be misconstrued to be an [OIII] emitter at 1.5<z<2.7. These catastrophic failures can be overcome by joint fitting of the multi-wavelength photometry and grism spectra necessitating joint analysis of imaging and spectroscopic data. Alternatively, the photometric information, particularly the isophotal area, can be used as priors to the spectroscopic fits, requiring ground and space based photometry matched to the Euclid images (Figure 9).

b.  Providing high spatial resolution, optical/near-infrared Galactic extinction maps to correct the multi-wavelength photometry across LSST/Euclid/WFIRST. This in turn, requires deriving the Galactic dust emission properties at the highest spatial resolution possible with the existing data (~10s of arcsec with WISE 12 and 22 microns) and calibrating it against the colors derived from precise optical/NIR photometry of stars that can be derived from joint processing e.g. (Meisner & Finkbeiner 2014, Schlafly et al. 2017). Obtaining consistent deconfused stellar photometry from the u-band through the NIR therefore provides the longest lever arm for calibrating both the extinction law and the normalization.

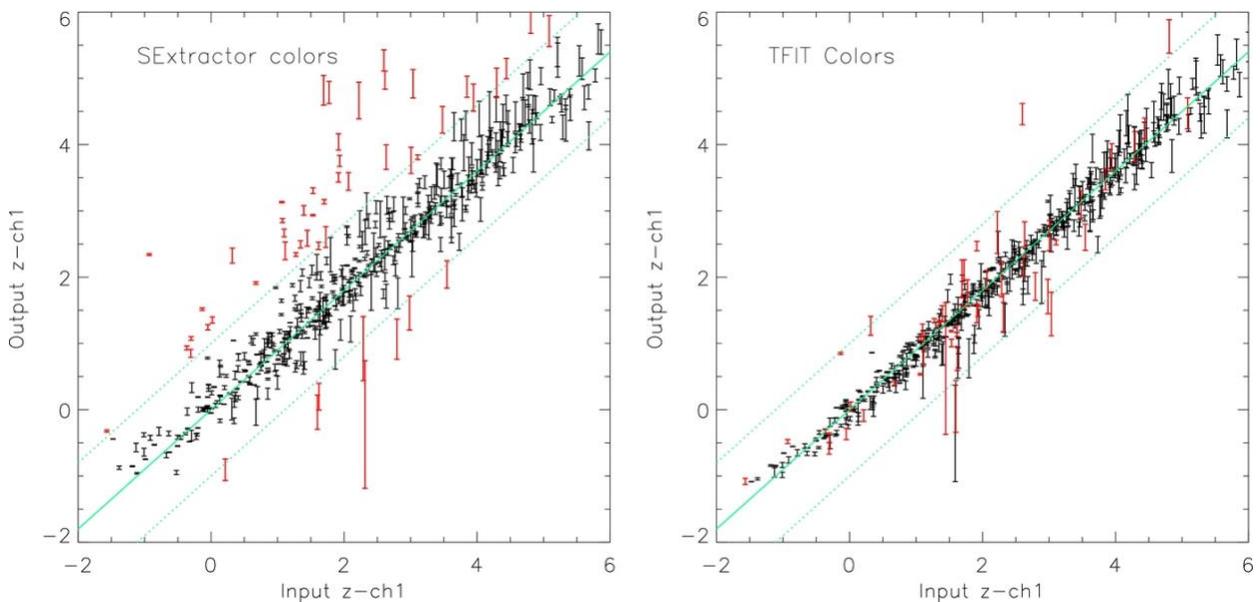

Figure 5: Simulation from Lee et al. (2012) showing improvement in the measured z-[3.6μm] (z-band is from Hubble/ACS while ch1 is Spitzer/IRAC channel1 at 3.6 microns) colors of objects when using high-resolution priors in position and morphology to extract photometry (right panel), compared to when using catalog matching (left panel). The red points are used to identify sources whose blind catalog photometry was biased by >1 mag. Priors clear reduce both the number of catastrophic outliers and the photometric scatter.

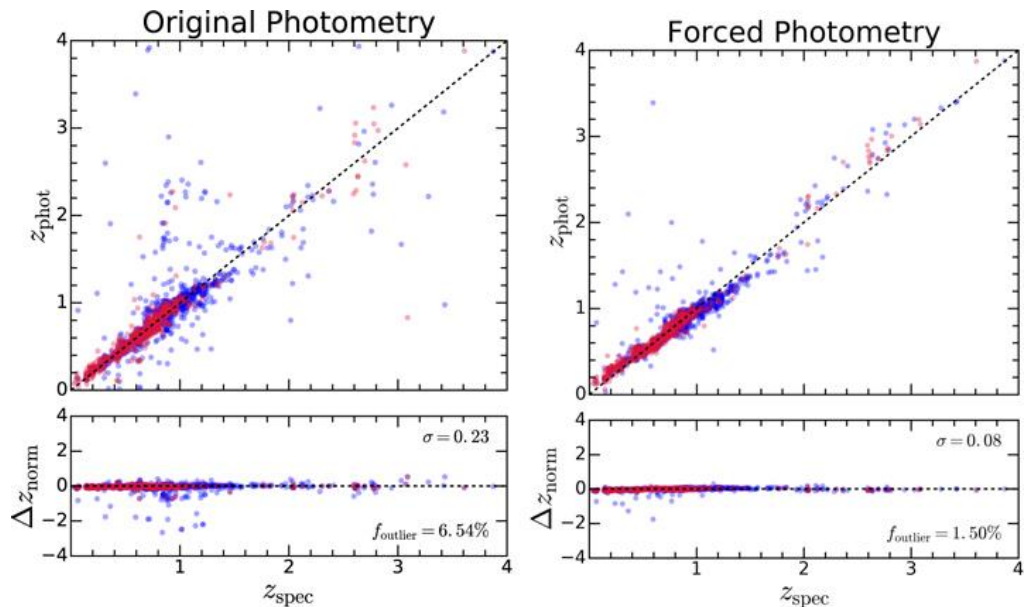

Figure 6: Comparison (from Nyland et al. 2017) between spectroscopic and photometric redshifts derived by using catalog matching (left) versus by doing forced photometry using positional priors (right panel). Red sources have high quality spectroscopic redshifts while blue sources have lower quality redshifts. The scatter in photometric redshifts is reduced by a factor of 3 while the outlier fraction has been reduced by a factor of more than 4. When combined with precision corrections to the photometry like field-dependent filter transmission and Galactic extinction, the errors in photometric redshifts can be dramatically mitigated at an even higher level through pixel-level joint processing of the data.

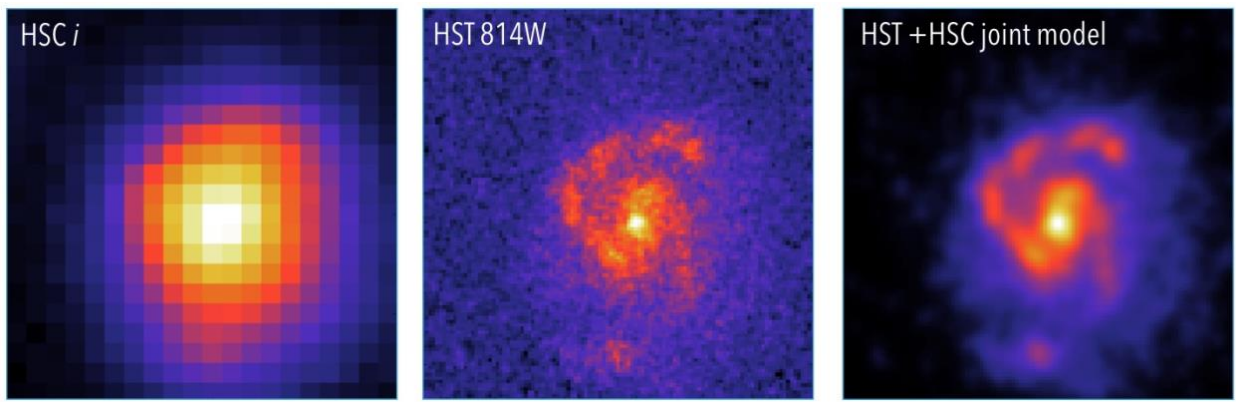

**Figure 7:** Joint modeling of ground- and space-based data. A spiral galaxies in the COSMOS field, observed by HSC *i*-band (*left*) and the Hubble ACS F814W (*center*), and the non-parametric joint model from SCARLET (*right*). The lower surface brightness limit of HSC allows fainter features in HST images to be modeled. Joint modeling is even more beneficial for the combination of LSST (comparable to HSC, $i \sim 27$) and the less sensitive Euclid (VIS < 25.2). Figure courtesy Rémy Joseph (Princeton).

2.    Improving the identification of Type Ia SNe and their classification based on the properties of the host galaxies in which they occur.

a) Type Ia SNe are standardizable candles, which have been used as distance rulers to identify the contribution of dark energy to the energy density of the universe. LSST will detect 1000s of Type Ia SNe across the entire sky out to z~1 while Euclid and WFIRST will, in their deep fields which are targeted repetitively, observe 100s of such SNe, particularly extending their redshift range out to z~2. The properties of Type Ia SNe have been found to be correlated with properties of their host galaxies (e.g. Sullivan et al. 2010). Events of the same light-curve shape and color are, on average, 0.08mag brighter in massive host galaxies (presumably metal-rich) and galaxies with low specific star formation rates (sSFR). In addition, the presence of bright host galaxies affects the precision with which the SN lightcurves can be extracted. Accounting for that requires the capability to identify the frames, in which the host galaxy light is not contaminated by the supernova. One then needs to make coadded images out of this subset of exposures and reconstruct multi-band photometry as well as stellar mass and star-formation rate, ideally by jointly modeling the host galaxy in space- and ground-based data.

b) Furthermore, the grism spectroscopic data taken by Euclid and WFIRST will be able to constrain the redshift of the hosts of SNe detected by LSST. However, if the hosts are not strong line emitters, their spectra may never be analyzed and have their redshifts extracted by the default spectroscopic pipelines. Joint processing provides the framework to decontaminate the spectra of sources of interest, optimally extract their spectra using the measured surface brightness profile of sources in the space-based imaging data (especially for resolved sources), and measuring their physical properties such as redshifts, dust attenuation, line fluxes.

3.  Strong lensing time delays in background transients and substructure in the dark matter distribution

The lensing of background sources by massive foreground galaxies is a probe of both dark matter properties in the lens and cosmological parameters, through the measurement of time-delays in the different images of the background source. The lensed images typically have too small an angular separation from the foreground source which results in blending in seeing-limited resolution data. Furthermore, it is the smallest separation configurations which provide the best tracers of the dark matter profile in the lens and can distinguish between cuspy and cored dark matter profiles when combined with spatially resolved kinematics. The extraction of multi-color information of both the lens and the background source as well as generating the light curve of the background source for differential time delays between the images, requires deconfused multi-epoch photometry. Joint analysis of space- and ground-based data helps disentangle the seeing-dependent contamination of the lensed source by the lensing source, enabling accurate redshifts, galaxy properties and time delays to be measured (Suyu et al. 2014; Birrer et al. 2017).

The number of strong lenses know at the present time is ~100, with the vast majority found through the SLACS survey. The Euclid wide area survey will detect a million strongly lensed galaxies with several tens of thousands of lensed QSOs (e.g. Boldrin et al. 2016 and Euclid Strong Lensing SWG White Paper). The temporal variability of these will only be monitored by LSST and other comparable ground-based telescopes, although the seeing will cause blending between the lensing galaxy and the multiply imaged QSOs or background galaxy. Using the Euclid image as a template to fit for the brightness of the lensing galaxy and the QSOs in the single-epoch LSST data will mitigate the effect of blending and yield accurate photometry through which the time delays can be measured.

Cosmology Requirements:

CoR1. Calibrated single frame data and metadata (e.g. seeing/PSF characteristics, epoch of observations, equinox of coordinate system, water vapor content, photometric conditions and instrument properties) from all participating projects.

CoR2. Concordance of astrometry and photometric standards across all wavelengths over the entire observed sky (Figure 8 and 9).

CoR3. Measurements of galaxy and point source properties using space and ground-based images and knowledge of the instruments e.g. filter transmission, PSF models using tools like TPHOT, TRACTOR and SCARLET.

CoR4. Joint survey processing pipeline that identifies overlapping subsets of sky area from all participating surveys, interfaces with joint measurement and analysis methods and generates catalog level output photometry, uncertainties and flag values.

CoR5. Custom extraction and coaddition of calibrated frames for SN cosmology and for time-domain studies. Uniform time standards (barycentric, geocentric, TAI etc.) across projects so that identification of transients by any one project is applicable to other projects and results in accurate light curves.

Cosmology Desirables:

CoR6. Integration of WISE and GALEX data with Euclid/LSST/WFIRST photometry which will increase the number of bands with flux densities for training spectral energy distributions of galaxies.

CoR7. Construction of a high spatial resolution Galactic extinction map using ancillary data from WISE, IRAS, Planck and HI maps combined with multi-wavelength stellar photometry from Euclid/LSST/WFIRST, after accounting for stellar proper motion between the surveys.

## 2.2. REIONIZATION AND GALAXY EVOLUTION

a. Detecting sources (both galaxies and active galactic nuclei) at the epoch of reionization and using the strength of Lyα emission to characterize the reionization history of the Universe over cosmic variance-unbiased volumes is one of the key goals of Euclid and WFIRST. The identification of candidate objects at z>6 will rely on deep, ground-based imaging with LSST or Subaru/HSC, which will enable accurate, multi-band color selection (e.g. V-I, i-z or z-y; Figure 5). Undertaking robust color selections by combining LSST and space-based data will require resolution-matched photometry and deblending of the contribution from nearby objects (e.g. Figure 10). This necessitates developing customized algorithms that will extract flux densities in the coadded images. Extracting and measuring the Lyα emission of these sources from the grism surveys will also require specialized, joint processing. The Euclid pipelines will generally perform automated spectral extraction of sources which are in the input imaging catalog, i.e. brighter than ~24 or 24.5 AB mag while typical z>6 galaxies will be much fainter unless they are lensed, which also introduces source confusion. The LSST astrometric images will need to be aligned with the Euclid/WFIRST grism images, and all individual contaminating sources in the 2D dispersed images subtracted, before the Lyα flux from the source of interest can be identified.

b.     For galaxies at lower redshifts, the grism spectra from Euclid/WFIRST will supersede, in most cases, the photometric redshifts derived from LSST data alone (see Figure 13 for example). Furthermore, their emission line properties will require careful flux calibration which will have to be derived after combining the grism spectra from different position angles (PA) – since the morphology of the source in the cross-dispersion direction is different, each of these PA's will have to be extracted, deblended and calibrated against the segmentation map derived from the imaging data, per position angle of observation. Such a procedure has been implemented before for the HST-3D survey (e.g. Momcheva et al. 2016) but the level of confusion is expected to be much higher, while the precision required for line flux estimation in Euclid/WFIRST is better than 10%.

c.     Studying the morphologies of LSST color-selected galaxies requires identifying objects based on a combination of LSST/Euclid/WFIRST colors, and extracting surface brightness profiles and morphology metrics from Euclid-VIS or WFIRST data (~0.1" resolution). In particular, quantitative morphology metrics will need segmentation maps to be matched which requires matching LSST and space-based data on a common pixel grid and/or comparing sensitivity-dependent morphological differences between Euclid and WFIRST (e.g. Figure 2).

Galaxy Evolution Requirements:

GER1. Matching LSST imaging data and Euclid/WFIRST grism data on a common astrometric reference scale.

GER2. Enabling the PSF-matched extraction of detection limits from optical imaging data to ensure robust Lyman-break color selection of objects by combining Euclid/WFIRST and LSST.

GER3. On-demand spectral decontamination of Euclid/WFIRST grism data.

GER4. Enabling the capability to run user-defined tools (such as morphology fitting) directly on pixel-level imaging data across multiple wavelengths for subsets of color-selected objects.

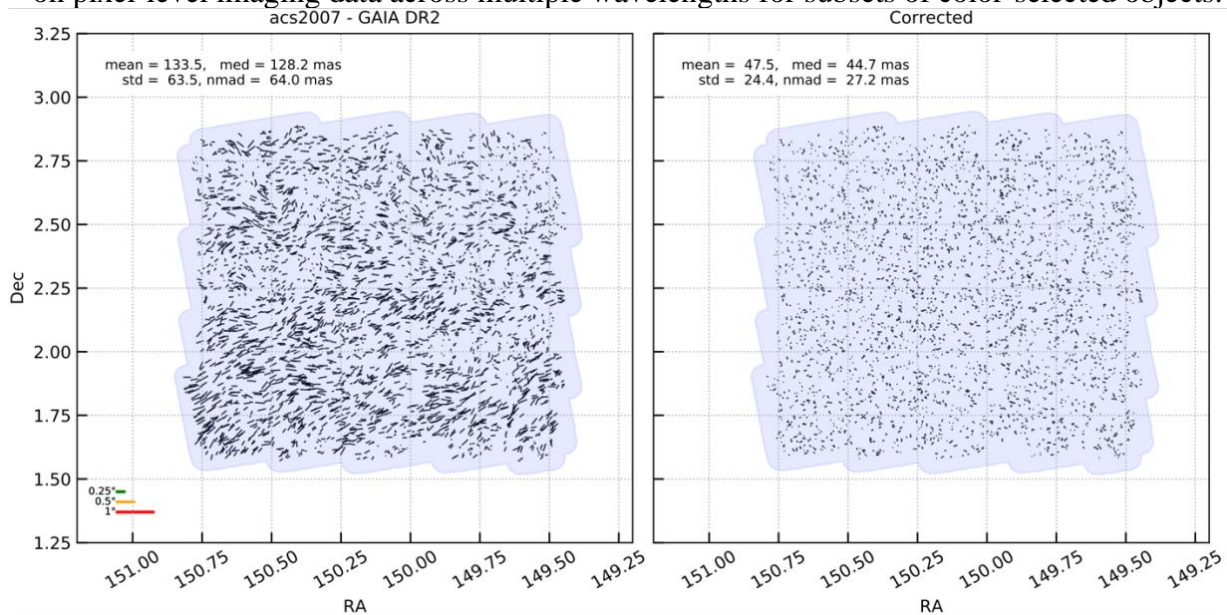

Figure 8: Figure highlighting the need for precise cross-project astrometry. The left panel shows astrometric distortions between Hubble ACS and Gaia DR2 in the COSMOS fields while the right panel shows the astrometric offset after corrections have been made. The reduction in astrometric scatter is a factor of ~3. Such statistical offsets in positions result in imprecise results for joint photometry of sources, typically underestimating the photometry of the sources as shown in Figure 9.

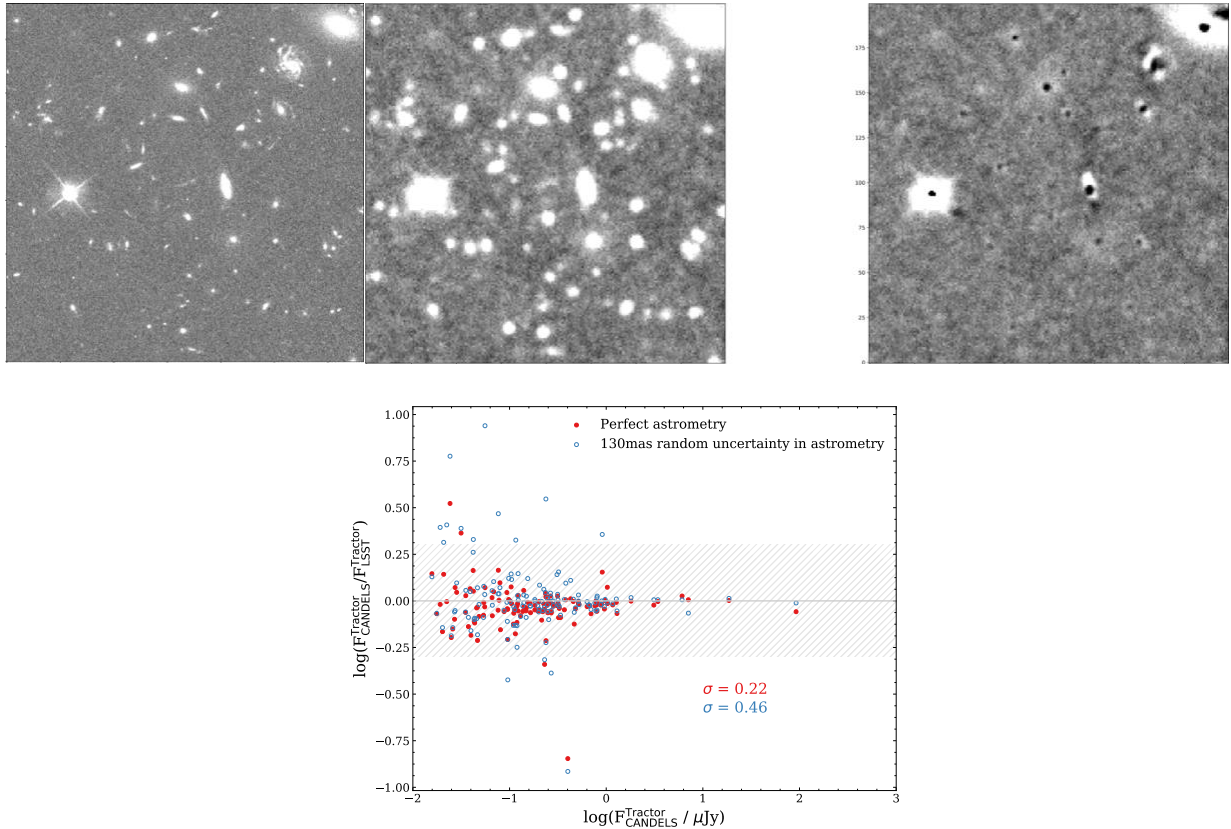

Figure 9: Left panel showing a high resolution WFIRST-quality R062 (optical filter) image derived using HST data from CANDELS (Grogin et al. 2011; Koekemoer et al. 2011). The two panels on the right show the equivalent LSST r-band image and the residuals after doing prior based photometry with TRACTOR. A comparison of the resultant photometry is shown in the lower panel, where the x-axis is the CANDELS photometry and the y-axis is the log of the ratio between the CANDELS and deblended TRACTOR photometry (hatched band shows a factor of ±2 difference). The red points show the photometric scatter with no astrometric error while the faint blue circles show the scatter when an arbitrary 1sigma astrometric scatter of 130mas is introduced on a source by source basis. The standard deviation quoted is the fractional error in the extracted flux densities; the standard deviation increases by a factor of ~2 due to astrometric inaccuracy. This figure serves to highlight the importance of accurate astrometry and its importance on photometric scatter.

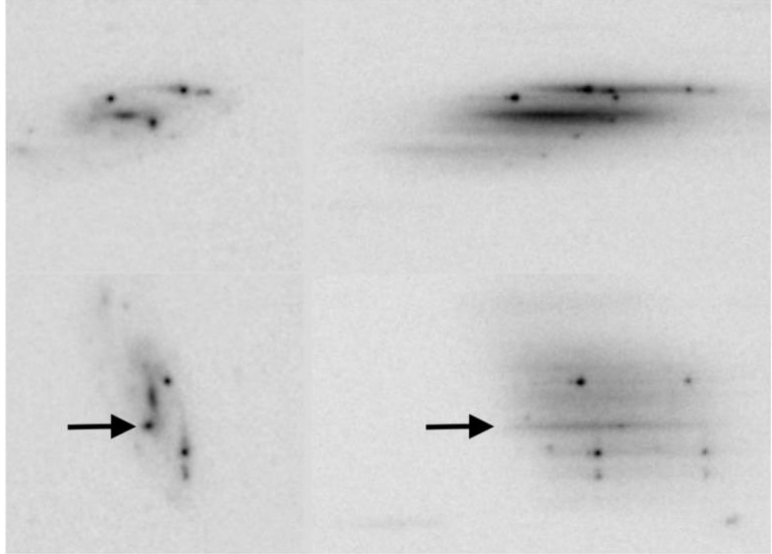

Figure 10: Illustration of spectral confusion. The left panels show the Hubble near-infrared images of a nearby resolved galaxy in two position angles rotated by 90 degrees. The right panels show their correspondingly dispersed grism spectra. The lower panel has an arrow marked at the location of a high-z Lyman-break galaxy whose spectrum is totally contaminated by the spectrum of the foreground galaxy in the upper panel. Joint pixel level processing will allow users to leverage containerized software tools to identify such sources using color-selection in the space imaging datasets and undertake customized extraction in the corresponding grism datasets of such sources.

## 2.3.   MICROLENSING AND THE SEARCH FOR PLANETS, SUPERSTELLAR MASS BLACK HOLES

Microlensing of background stars towards the direction of the bulge has been demonstrated to be a powerful technique to characterize the population of super-Earths and provide constraints on the number density of free-floating black holes in our Galaxy. Bulge fields due to their high stellar density will be affected by source confusion, particularly in ground-based seeing limited datasets, such that microlensing events (seen by LSST) will only be detected at a sensitivity level that is 3-4 mags brighter than the shot-noise limit of individual images. By using space-based data from Euclid/WFIRST to undertake a joint fit for all the non-varying stars, one can detect much fainter microlensing events with LSST, particularly in fields where the space surveys do not have a large number of repeats. This in turn, will enable the detection of lower mass planets in fainter, more distant stars and free-floating super-stellar mass black holes. Robust identification of the mass of the planet will also require an accurate 3D extinction map towards the stars which will be developed as part of the earlier requirements, by combining WFIRST, Planck and HI maps.

Microlensing requirements:

MLR1. Photometry is derived by deconfusing single-epoch LSST frames taking seeing variations into account, while leveraging deep space-based data priors (e.g. positions and brightness) of stars and Gaia derived proper motion solutions.

MLR2. 3D Galactic extinction map to determine the intrinsic luminosities of the stars.

## 2.4.  MOVING OBJECTS: STELLAR PROPER MOTIONS AND SOLAR SYSTEM OBJECTS

### 2.4.1.  STELLAR PROPER MOTIONS

a.    Gaia will provide parallaxes and proper motions of stars to faint levels (G~21 AB mag). Stars would have moved, albeit slightly, between the epoch of Gaia data taking and the epoch at which the Euclid/LSST/WFIRST frames are taken (Figure 11). By allowing true cross-identifications (not just static positional cross-matches) across epochs, for stars detected by Gaia, taking into account their proper motions, the pixel-level combination of LSST, Euclid and WFIRST surveys will: 1) greatly alleviates source confusion towards the Galactic Plane, towards the inner Galaxy and in external galaxies, enabling more accurate photometry and proper motion to be extracted and 2) provide absolute optical-to-IR luminosities, which are very nearly absolute bolometric luminosities for most objects, since there is little flux shortward or longward of these wavelength regions covered by these surveys.

While Euclid only intends to observe the high Galactic latitude sky as part of its primary survey, LSST and WFIRST will undertake observations of the inner Galaxy. Thus, deblending of LSST data using WFIRST priors as shown in Figure 9 will allow the reconstruction of the multi-scale structure of the Galaxy from a few pc (Solar Neighborhood) to 10s of kpc in the Plane and the Bulge. Furthermore, when combined with the effective temperature ($T_{eff}$) from either the Gaia spectroscopy or from our reassessment given the shape of the optical/NIR spectral energy distribution, the radius for a huge sample of stars can be derived using the Stefan-Boltzmann law. These calculated radii across a wide swath of the H-R diagram provides an independent check of radius for those objects with interferometrically measured radii. We will be able to confirm the scale height distribution as a function of spectral type, as theoretically predicted by Ryan et al. (2017), according to which the scale height has a minimum for L/T transition objects and then increases for M-type dwarfs, consistent with cooling models. At distances of hundreds of kpc (Local group dwarf galaxies), the Euclid/WFIRST priors will greatly improve proper motion measurements of bright giants in the dwarf galaxies allowing inferences on the distribution of dark matter in these galaxies.

b.    A similar joint photometric analysis in globular clusters would result in better sampling of their spectral energy distribution and separation of stellar populations and by metallicity, age and binding energy. The addition of high-resolution space-based imaging will allow these studies to probe much further inside the globular cluster which in turn will provide greater insights on the star-formation history of globular clusters and reveal the role of globular clusters in the early star-formation history of galaxies and their contribution to the reionization process.

c.    The measurement of halo kinematics and the detection of Galactic substructure (streams, dwarf galaxies, etc.; Figure 12) will be tremendously benefited by LSST proper motion measurements at distances much larger than Gaia can reach. By itself, LSST will be able to measure proper motions to 1 mas/yr (50 km/s at 10 kpc) down to r ~ 24, and 0.2 mas/yr (50 km/s at 100 kpc) to r ~ 21. Euclid in its wide-area survey and WFIRST through GO surveys of streams will measure fewer but potentially more accurate positions for all stars in the fields of overlap, thereby providing an accurate reference frame for comparison and providing a space-based time baseline that is comparable to that of LSST. Furthermore, by using the space-borne platforms to establish an astrometric grid, the LSST sources can also be combined with 2MASS to establish a longer time baseline spanning several decades.

Galactic Science Processing Requirements:

GSR1. Establishing a common time baseline for the frames to derive improved proper motion measurements.

GSR2. High source-density deblending and single-epoch PSF-fit photometry for the entire survey.

GSR3. Galactic extinction correction, including accounting for variation in the extinction law, to derive intrinsic luminosities.

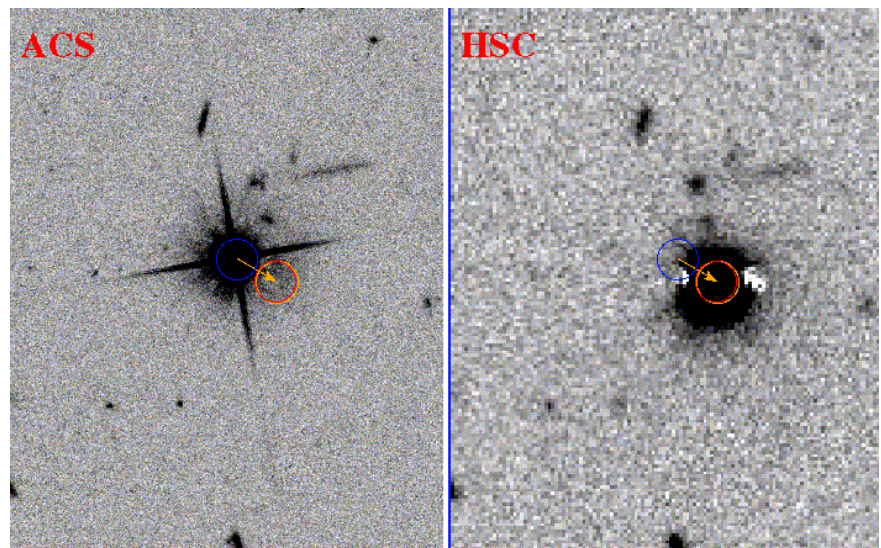

Figure 11: Illustration of proper motion derivation using two epochs of observations from space (left: Hubble/ACS, epoch of 2004.58) and the ground (right: Subaru/HSC, epoch of 2015.06) which joint pixel level processing will enable. The 17.4 mag star shown has a proper motion of 200mas/yr (as determined by Gaia). The position in the Gaia reference epoch is shown as orange circle, the predicted or proper-moved position at the ACS and the HSC epochs are shown in blue and red, respectively. With an expected 1mas/year astrometric precision, and better handling of confusion by nearby sources, JSP will enable both a verification of Gaia proper motions and measurement of proper motions for much fainter objects over a longer time baseline.

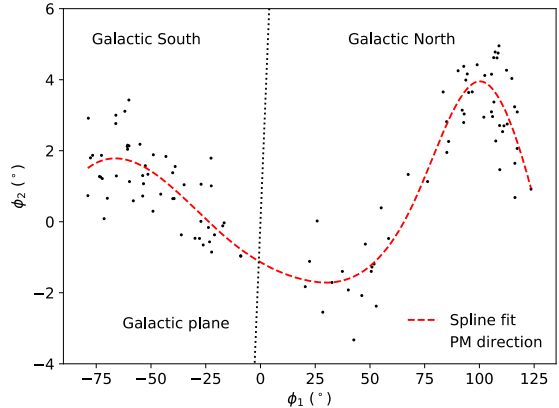

Figure 12: The distribution and proper motions of RR Lyrae over length of the Orphan Stream, in a coordinate system roughly aligned with the stream. The gray arrows show the directions of Gaia-measured proper motions, averaged over 25 degree intervals. They reveal that, while the stars in the northern part of the stream are moving along the path of the stream as expected, the stars in the southern portion of the stream (phi1 < 25 degrees) are moving nearly perpendicularly to the path of the stream. This surprising ~0.5 mas/yr sideways motion is consistent with the fly-by of the Large Magellanic Cloud in the south, if the LMC has a total mass about 10% that of the Milky Way. JSP will enable astrometric measurements of a larger set of fainter stars in such streams (Adapted from Erkal et al. 2019, MNRAS, submitted). Although the single epoch astrometry achievable will be ~10-20mas, by averaging over a large set of stars fainter than those observed by Gaia over a 10 year Euclid+LSST baseline, bulk proper motions as small as those seen here can be achieved.

### 2.4.2. SOLAR SYSTEM OBJECTS

The combination of multi-epoch imaging from LSST and space-resolution imaging promises to have a significant impact on solar system science. It is unlikely that Euclid will image the Galactic Plane or Ecliptic Plane until end of prime emission (2028-), after which dedicated surveys may be undertaken. However, WFIRST will undertake dedicated surveys of the inner bulge, and the Euclid high latitude survey will generate an important data set that can be searched for observations of objects at these locations such as near-Earth and Mars-crossing asteroids. The main merits of combining these data sets lie in 1) reducing confusion noise in the images by fitting for and subtracting known sources in the field providing better sensitivity to moving objects and 2) providing a long time baseline (> 10 yrs) over which to measure the positions of solar system object, improving orbital solutions.

Since the spectra of SSOs at these wavelengths mostly follow a solar reflection spectrum, the optical bands of the space missions and LSST provide the greatest potential for exploring these populations. In the case of small Solar System Objects (e.g., asteroids, comets, Kuiper Belt objects, etc.) the main science driver is astrometric pre-covery (orbit solution at epochs before detection) and recovery (orbit solution for future epochs). These data allow for improved constraints on the orbits of the targets, and improve predictions of their future positions. For objects that come close to the Earth, these further constraints can be used to rule out Earth-impacting solutions, while for spacecraft mission targets extended orbital arcs allow for simpler mission planning. Pre-covery and recovery can be achieved by non-sidereal stacking of all relevant single-epoch frames from LSST, Euclid and WFIRST along the path of motion of the object, or by applying a source identification tool such as MOST (Moving Object Search Tool)₁ available at IRSA (Infrared Processing Analysis Center) to the single-frame images. Furthermore, since source confusion will be significant in the LSST data close to the ecliptic and Galactic plane, the stacking can be done on images with all fixed sources subtracted using priors from the space-based data which dramatically reduce the confusion noise. Since the single epoch magnitudes of LSST and of the Euclid survey are comparable, every single Euclid source can be fit for and subtracted in the LSST images leaving only the time variable and moving objects.

The combined LSST/Euclid and WFIRST photometric data points also allow to set broad constraints on the object chemical composition since the albedo is proportional to its composition. Photometric colors have been successfully used to provide broad taxonomic classifications of asteroids in the optical (Carvano et al. 2010, A&A, 510, A43) and near infrared (Popescu et al. 2018, A&A, 617, A12), which could be expanded upon by combining LSST, Euclid, and WFIRST photometry.

Finally, with sufficient data and a sufficiently long baseline, it is possible to reconstruct the shape of small SSOs using sparsely sampled light curve photometry using light curve inversion techniques (e.g. the Database of Asteroids Models from Inversion Techniques [DAMIT]; Durech et al. 2010, A&A, 513, A46; Durech et al. 2018, A&A, 617, A57). By fitting an assumed color to handle the difference between the observed optical and NIR albedo, coarse SEDs can be probed as well., This requires accurate absolute calibration across the photometric bands.

Solar System Objects Processing Requirements:
SSO1. single-epoch frames, both with and without source subtraction with non-sidereal stacking capability;
SSO2. absolute astrometric precision of 0.1 arcsec which needs to be achieved after taking into account the proper motions of stars between the different epochs of LSST/Euclid/WFIRST observations;
SSO3. absolute cross-band photometric precision, with 1 sigma of 0.05 mag.

# 3. Technical Requirements for Joint Survey Processing

The scientific studies described in section 2 require a data processing system that ingests images and spectra from multiple surveys (Tier 0); consolidates their astrometric and photometric definitions in a concordance frame (Tier 1); performs joint analysis on the concordance data (Tier 2); and exposes the concordance data and JSP catalogs to the astronomical community for follow-up studies (Tier 3).

We should emphasize that none of the tasks we outline below are intended to duplicate work already performed by the individual survey projects (e.g. instrumental calibrations, or dedicated analysis tools). The goal here is to leverage this work to construct a high-fidelity concordance data set and provide online application program interfaces (APIs) for the astronomical community to carry out novel analyses on the concordance dataset. Furthermore, while we hope for a formal data sharing agreement between the surveys, all efforts described here can be executed with public data sets that are going to be released by each of the survey projects, as per their data release schedule.

## Tier 0: Access to Survey Data

The fundamental requirement for any JSP system is to provide access to the individual survey data when needed.

- LSST: 70 Petabytes ($10_{15}$) of calibrated data, corresponding to 7 Petabytes/year starting 2022.
- Euclid: 2 Petabytes of calibrated data over 6 years, corresponding to 0.3 PB per year starting 2022.
- WFIRST = 18 Petabytes corresponding to 3 Petabyte/year starting 2026 for the six-year prime mission.

We therefore estimate the volume of data to be accessed for JSP to about 100 PetaBytes, which after concordance astrometry and calibration, will be doubled in volume. That estimate does not include the volume of the concordance stacked data set product which would be ~10PB in all.

**Design consideration:** In a conventional system, it would be desirable for the data to be located at a single data processing center, but the requirement can also be met with a distributed archive with the data centers for each project storing the data products that JSP will access. The alternative solution can leverage investments already made in the data centers of the survey projects and avoids the construction of a massive data center that essentially duplicates those data. Keeping the individual survey data with the projects would also transparently benefit from any incremental recalibration of the data as routinely performed by the projects.

For a distributed system, JSP Tier 0 requires

- A **centralized mechanism** to dispatch queries, typically for some or all of the available data in a spatial region of the sky, to the individual survey projects.
- A **transfer protocol** to interface with the survey data centers and transfer the data (calibrated images/spectra and metadata) to a computing center for joint processing.

In section 4.3 below, we estimate that peak data transfer rates can be of the order of 1 PB/day, with large-scale requests preferably being scheduled with the projects. We note that current multi-stream transfer rates between domestic data processing centers are ~10 Gbps, so there is a clear need for improvement or optimization to bridge the ~1 order of magnitude shortfall. As fast network connection demands are common for advanced data analysis systems across the sciences, we consider our approach as an example use case that would benefit from dedicated investment in the U.S. network infrastructure, and may well be met within the next few years. Alternatively, it is conceivable to push some of the low-level analysis code to the location of the data to reduce the network loads.

## Tier 1: Concordance Frame

The next requirement for JSP is that retrieved data from the projects need to be brought in a concordance frame, in particular in terms of astrometric and photometric conventions, but also the time of observations. This will be critical for downstream analyses, as we demonstrate in Figure 8 and 9, and is not in scope for each survey project, for which internally consistent conventions are normally sufficient. In Figure 14 for instance, we illustrate the impact of photometric discordance on the derivation of photometric redshifts of galaxies. JSP aims to ensure photometric concordance between surveys is at a level that is at least as precise as the photometric concordance within a survey.

In detail, JSP Tier 1 requires:

- **Astrometry standardization** with cross-correlation to Gaia proper motions. This task entails determining stellar proper motions from Gaia and possibly a deeper survey, the epochs at which data were taken, the estimated positions of stars at those epochs, and deriving the astrometric correction for each data frame.
- **Photometric standardization** to ensure consistent corrections for airmass, extinction, moonglow, sky background, etc and their application to each data frame.

**Design consideration:** Given the high density of reference stars for the astrometric and potentially photometric re-calibrations, it is advantageous to hold Gaia and external stellar catalogs at computing site for JSP. All other meta-data can be retrieved from the project data centers through JSP Tier 0.

An important aspect of Tier 1 is the construction of a unified testing and quality assessment framework that demonstrates that the tight allowances for astrometric and photometric cross-survey calibrations have been achieved.

## Tier 2: Joint Analysis

Once all data of a spatial region of the sky is transferred to the computing site, the concordance frame is established and the data are jointly analyzed by carrying out these steps:

- **Detecting sources** from the matched data.
- **Star—galaxy separation** exploiting the information from all surveys.
- **Simultaneous fitting** of point sources and galaxies, either by utilizing space-based images as priors of morphologies and positions, or by jointly fitting of all available exposures. This step includes making decisions on the optimal extraction technique based on the character of the sources (in particular which of them are point or transient sources) as well as accounting for differences in photometric system, PSF shape, and the treatment of overlaps of neighboring objects.
- **Catalog generation** and storage containing a canonical set of measurements (position, flux in every band, morphological fits, estimates of variability, processing flags) as well as by-products such as upper limits and residual images. This catalog should include a master detection catalog that uniquely identifies sources across the individual surveys.

The optimal methods to perform cross-band detection, star—galaxy separation, and simultaneous fitting are being actively developed, in particular by the individual surveys. We envision to start with a conventional approach, where detection of sources is largely based on high-resolution space-based imaging and stellar catalogs are pre-established from proper motion or variability measurements by Gaia or LSST. This baseline analysis pipeline can be extended when survey-specific refinements are made by the individual survey projects. In addition, leveraging cross-survey information to advance substantially object detection, characterization, and de-confusion can only be achieved through dedicated method development in the context of JSP Tier 2.

### Tier 3: Science Platform

The generation of the concordance catalog and data products will immediately serve many science cases (as outlined in section 2 above). But even more analyses can be performed when the data and software elements of JSP are made publicly available through web interfaces and APIs.

The key aspect of the Science Platform which containerizes various software packages that are utilized upstream of the concordance product generation, is that it fully leverages developments from Tiers 0 to 2 to carry out cutting-edge investigations that will only emerge over the next several years.

To support common patterns of special-purpose queries, the JSP Science Platform needs to support these tasks:

- **Generation of coadded images** from subsets of the data. This task includes the ability to select based on time, or seeing or photometric conditions or source variability properties, and the possibility to coadd sources with non-sidereal motion.
- **Filtering images with custom kernels** e.g. to extract diffuse source contribution from tidal streams and dwarf galaxies.
- **Fake object insertion** to test their recovery by the JSP system.
- **Spectral extraction** of the grism data of Euclid and WFIRST to determine the 1D spectrum of a source of interest, while decontaminating it from neighboring sources (e.g Figure 10 and 13).

In addition to a central JSP Science Platform, we strongly endorse the public release of the JSP code base and documentation such that our approach of accessing, standardizing, and analyzing data from multiple surveys can be applied to surveys outside of our immediate scope. Examples that are of considerable scientific interest include (1) joint analysis of data from WISE and GALEX, which allows FUV, NUV and mid-IR (3-22 µm) flux densities to be extracted, and (2) extending the high-resolution Galactic extinction calculator by matching HI, WISE and IRAS/Planck data, as well as future relevant surveys.

Table 2 shows a summary of the tiers of data processing tasks and the resulting data products. Table 2 also shows the code of the science requirement from Section 2 that is satisfied by the tasks or products in those cells. An exhaustive formal treatment of requirement flow-down will be undertaken when JSP work is approved to proceed, but this Table was sufficient to support the scoping work.

Table 2: Tiers of Data and Software Products from JSP

| | | | | |
|---|---|---|---|---|
| **Tier 0** | Single epoch frames and metadata [CoR1, GSR1, SSO1] | Single-frame astrometric solution [CoR1, SSO2] | Color/location dependent PSFs [CoR3, GER2] | Photometric calibration parameters [CoR1] |
| **Tier 1** | Concordance solutions for astrometry and photometry [CoR2, GSR1] | Optimal coadds, noise and coverage maps [CoR4, GER2] | Interface to external datasets (e.g. Gaia) and PSFs for forced photometry [CoR6] | Astrometrically aligned spectral products [GER1] |
| **Tier 2** | Joint photometry tools, catalogs, residual maps [CoR3] | High resolution Galactic extinction maps [CoR7, MLR2, GSR3] | Point source fitting tool [MLR1, GSR2] | |
| **Tier 3** | Science platform [GER4] | Custom image coaddition tool [CoR5, SSO1] | Cross-project fake source injection software [CoR4] | Spectral extraction and decontamination tool [GER3] |

# 4. Architecture

## 4.1. Vision

In response to the scientific opportunities, we have developed and scoped a vision for a tiered approach to JSP that balances the need for a stable, widely applicable global solution with the need for special-purpose processing. The JSP effort would release to the research community a data product that captures the astrometric and photometric concordance between the surveys, and offers pixel-matched resampled images and resolution-matched images. The access to the data product would be published as a containerized science platform that allows the user to leverage that concordance for investigations involving pixel-level cross-survey manipulation in pursuit of new science goals.

The concordance product would feed directly into the Dark Energy analysis, which is expected to be run only once for each major new release of data by any of the survey projects. Customized processing, whether targeting maximal sky coverage or focused on certain areas, would also start from the concordance product and operate on subsets of the data archives. Clearly, it is most cost effective to develop these capabilities as a coherent, robust service for all users by building on the insights and expertise of the three survey projects working closely with a dedicated JSP Core Team.

## 4.2. Schedule Constraints

All dates in this section are anticipated as of January 2019. ESA has baselined the launch of Euclid in June 2022 with a 6 month in-orbit checkout and cruise phase followed by a ~6-year survey; sky area coverage increases linearly with time, and there are no revisits over the 15000 deg$_2$ wide area survey. The deep survey and calibration fields (40 deg$_2$) will have near-monthly revisits but those are a tiny fraction of the total data volume. Euclid does not issue transient alerts, although sources which are varying in the deep fields may be studied by consortium members. LSST starts "full survey operations" in October 2022 and reaches Euclid depth in less than a year, with  depth increasing with time but not the fractional area overlap unless LSST commits to a significant survey in the accessible regions of the Northern Hemisphere as well. The current launch date for WFIRST is September 2025. In view of the LSST and Euclid calendars, we are targeting 1 September 2022 for system version 1 to be operational, so it is ready for ingesting the first public data releases from Euclid and LSST. It will continue to be developed for 2 more years (to September 2024) to complete functionalities and to integrate adjustments for the real data streams from LSST and Euclid. Further development work will continue after the WFIRST observations are initiated. Estimating 3 years to build the first working version of the system, we then need to start the development of the databases and JSP infrastructure at the start of FY2020 (October 2019).

## 4.3. Data Volume and Processing Constraints

The total data volume that we will need to have accessible at the end of the projects is about 100 Petabytes. Since LSST is a ten-year project, Euclid and WFIRST six-year projects, we anticipate that the data volume will build up at the rate of 10 Petabytes/year. If co-added images are the only data required, the total volume remains ~10PB. Of this, an ambitious but not extreme user may want to access a modest fraction, about 1 PB, corresponding to a few thousand square degrees of sky of co-added images, or a few hundred square degrees of full history images. It is reasonable to anticipate one such query per week, and anticipate that it represents the $95_{th}$ percentile in its ambition, i.e. data volume per query.  Experience with IRSA and other archives shows that the number of queries follows a power-law function of the data volume per query (Figure 15), with the ambitious query above and larger ones accounting for 90% of the data volume downloaded.  We therefore extrapolate from this ambitious request to a total of hundreds of significant (>1TB) data requests per year, with a total data transfer volume exceeding 1PB/day.

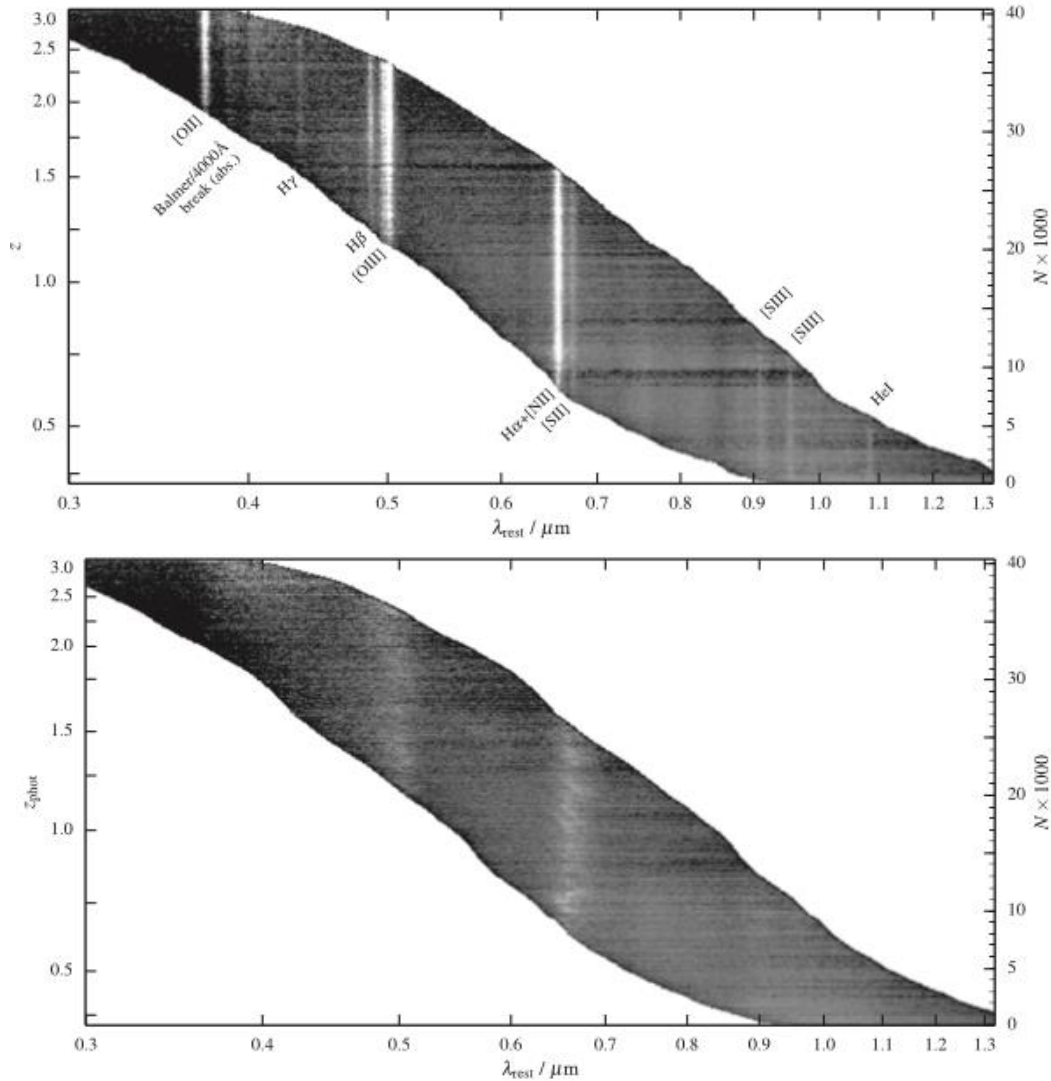

Figure 13: The top panel shows the sample of rest-frame spectra from 3D-HST from Momcheva et al. 2016. Each row shows the median of 100 individual 1D spectra, sorted by redshift. Strong absorption and emission lines that move through the WFC3/G141 passband are indicated. The lower panel shows the same figure but using photometric redshifts to order the spectra. The use of the grism spectra naturally improves the precision of redshift estimates. However, if one were to tease out faint absorption features in stacked spectra of galaxies selected by a certain color criterion, or obtain precise line flux estimates for galaxies consistent with their extended morphologies, one would have to decontaminate and extract the spectra, cross calibrate them with the imaging data after matching the isophotes and then coadd them. The cross calibration between the imaging and spectroscopic data is particularly relevant for baryon acoustic oscillation (BAO) studies but impacts all science areas that leverage the spectroscopy.

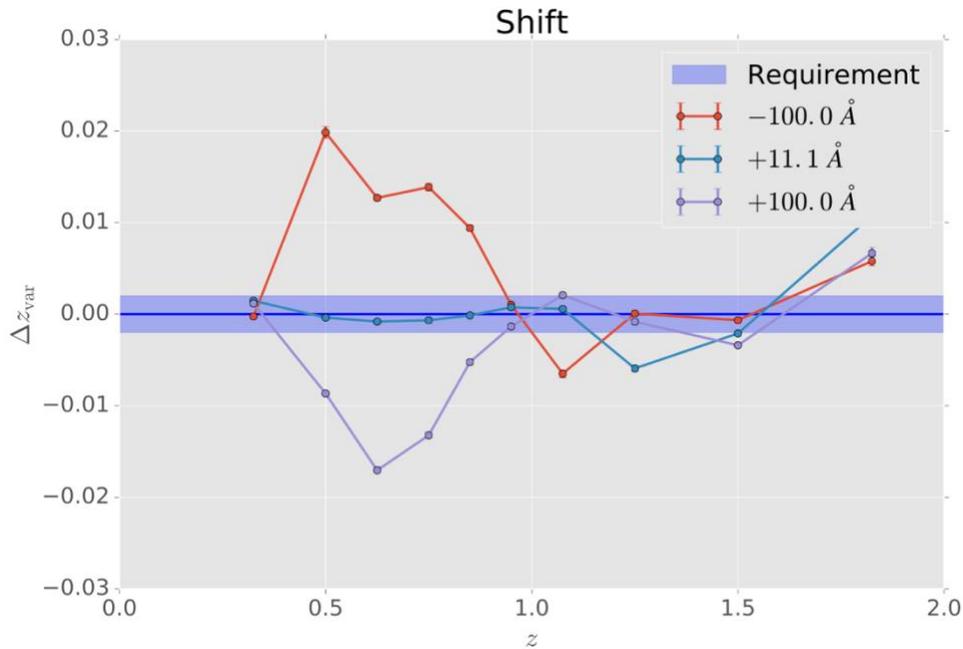

Figure 14: Plot illustrating the error in photometric redshift (y-axis) as a result of inaccurate calibration in a single band. A 10% photometric bias in a single band (R-band in this case) results in a shift in the median z of 0.02 (J. Coupon et al., Euclid internal photometric redshift report). The red and purple lines represent the direction of the bias i.e. brighter flux densities result in lower derived redshifts. For precise measurements of cosmological parameters, we need to eliminate all such biases ensuring that the calibration between projects (e.g. Euclid and LSST) is at the same level of accuracy and precision as the calibration within a project.

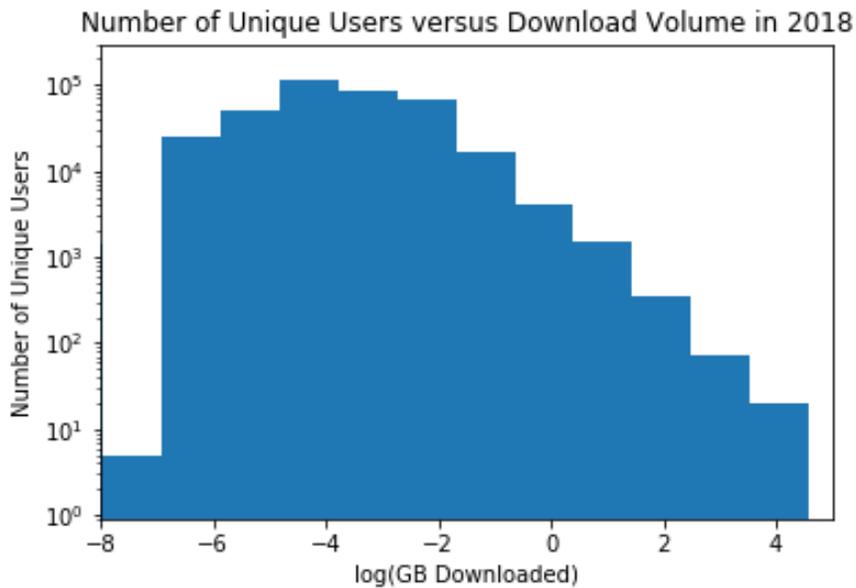

Figure 15: Plot showing the amount of data downloaded versus the number of unique users. The typical queries of archives tend to be small interactive queries while the data volume is dominated by rare large queries.

One of the key data volume and infrastructure drivers is whether there is a need to perform joint pixel processing on single-epoch calibrated frames or whether it is adequate for this processing to run on coadded mosaic frames, particularly from LSST, which would reduce the computing and storage requirements by one to two orders of magnitude. Section 2.1, 2.3 and 2.4 have previously outlined the scientific motivation for working on single-epoch calibrated frames for temporally varying phenomena. However, for the cosmology measurements in Section 2.1, we have not specified if single-epoch processing is necessary. Photometric measurements based on coadds are demonstrably less precise than those made by simultaneous modeling all epochs ("multi-epoch fitting"). As an example we show analysis of seeing-limited, ground-based data from Subaru/HSC and PanSTARRS (Figure 17 and 18). Our analysis finds photometric uncertainties of relatively compact sources to be about 15% worse when using coadds instead of a multi-epoch analysis. The seeing and its variability in the Subaru dataset are somewhat better to expectations for LSST (median~0.9" FWHM as per Appendix D of Ivezic et al. 2018). Furthermore, as part of our investigation we also contribute to an ongoing study (Armstrong et al., in prep) which finds that the reduction in measurement precision is proportional to the amount of variation in the ground-based seeing. This analytic derivation is confirmed by dedicated image simulations and implies that for the expected seeing distribution of LSST the loss in precision of the point-source photometry is of order 10%, and less for extended sources (see also Figure 17).

In addition, most of the morphological information for the JSP analysis can be extracted from the high-resolution space-based data of Euclid and WFIRST, rendering the task more amenable to use of coadds, with the aforementioned gains in computational efficiency. Thus the result of our investigation seems to show that for the overall concordance catalogs, it is adequate to work with the coadded data.

Finally, in the case of variable and moving sources, starting with the more computationally manageable coadd analysis provides a highly significant morphological model of all stationary sources. In a second step one can then determine the properties of the variable sources in a multi-epoch analysis, while keeping the stationary objects fixed. This reduces the runtime of the processing and renders the overall model more robust against noise and potential image artifacts in single-epoch data. It would be highly desirable for the JSP framework to enable this second step for variable sources of interest, which is also consistent with the demands of other time-domain analyses (as laid out in Sections 2.3 and 2.4). The implication is thus that all single-epoch images need to be available, with a full data volume listed in Section 3, and that multi-epoch processing could be limited to variable/moving sources.

## 4.4. Computing Architecture

We use software containerization in order to facilitate the reproducibility of our processing, enable computational scalability and give the user access to the standard software tools developed either by the project or part of JSP (Figure 16). Containerization allows us to ensure that the exact same versions of software libraries and applications are used, regardless of the software configuration of the machines that perform the computational work. Specifically, IPAC has adopted Singularity, a containerization technology developed at Lawrence Berkeley National Laboratory, which is widely used in supercomputing centers due to its compatibility with message passing interface (MPI), batch systems, GPU-accelerated code, and specialized communications protocols, such as InfiniBand. We store all of our analysis software in a single immutable singularity image file, which can be easily distributed to other machines. The software allows external users to access and visualize and manipulate the different imaging and spectroscopic data products and catalogs, as a function of time, sky coordinates, date of observation and wavelength band. For instance, a user may want to stack the individual frames in non-sidereal coordinates for which there will be standard software available. Multiple isolated container instances can run simultaneously on a single machine, which, for instance, allows multiple users to share a machine while exploring and manipulating data using Jupyter notebooks. This makes the software environment scalable where a user can install their preferred tools in their instance of the container and have it readily accessible for future initiations of their container. This dynamic architecture also allows for users to implement their "big data" tools to select parameters from the unified metadata and undertake correlation analyses between them on the unified catalog level data products that result from JSP.

We anticipate based on our preliminary TRACTOR runs on simulated Euclid/WFIRST and LSST data sets that the most basic joint survey cataloging effort on image coadds over 10 bands will take of order 30 million CPU hours. Monte Carlo simulations of the data sets for systematics, and crucially for completeness and reliability estimates would take 10 times longer than the all sky cataloging, namely about 300 million CPU hours. These Monte Carlo simulations are critical to characterize the joint survey catalogs properly, and are part of the JSP Core effort. These computational needs are best met at a single center with a robust infrastructure such as a national supercomputing center.

Examples of major bulk processing runs could require substantial resources as well. Extracting the microlensing light curves towards the dense Galactic bulge fields after de-confusion would take about 10 million CPU hours. Alternately, measuring the single-epoch photometry of all Euclid-detected lensed quasars in the 6-band LSST data with 200 epochs would take only about 60000 CPU hours. In addition, individual users who intend to manipulate the datasets will be requiring computational time. Assuming that 1000 astronomers per year intend to use the JSP datasets, we expect that about a billion CPU hours will be required from the different high-performance computing centers, the commercial cloud and institutional clusters. The previous largest allocation by the National Supercomputing Centers was $10^7$ hours for Planck. Clearly the computational resources required for access to, and working, with these datasets are so large that distributed resourcing is the preferred approach.

The hardware on which the containers can be run will be distributed. There will be limited computing resources at the data archives (typically a 200-core cluster) which will be ideal for interactive users who wish to initiate a container for a quick look at the data. The containers will also be portable to public-access high performance computing centers/open science grid or cloud service providers such as Amazon Web Services/Azure. We will leverage existing tools to monitor network capacity between the computing sites and the sites of data storage so a user is fully informed about the time it will take for their batch job to execute on these other sites rather than on the archive computing nodes. We will also build into the platform the ability to self-regulate rate of data requests to avoid overloading the data source archive or the network capacity. Enhancing those few critical network circuits will be the only infrastructure aspect relevant to the JSP enterprise, and our assumption is that it will not require JSP resources. We have demonstrated 10 Gbps networking between the IPAC archive and NERSC sites but that should be considered a lower limit based on the current firewalls and network switches. Over these networking speeds, processing a packet of data, spanning roughly a square arcminute in 2 bands (~10 GB) will take about 20s, if one requires the processing to include a customized coadd of single epoch LSST frames as well.

This hybrid architecture, distributing computation between the archives and shared HPC resources, will reduce the computing and disk space cost that will be associated with a particular user's science project. By having computational power located through high bandwidth connections to where the voluminous data are, and enabling manipulation of products served by the archive, passive archives will become flexible, interactive archives giving the capability for a user to share their code with other users and link them in publications – this enables astronomers to easily reproduce plots generated by others and increases the value of both the data and the resultant publications.

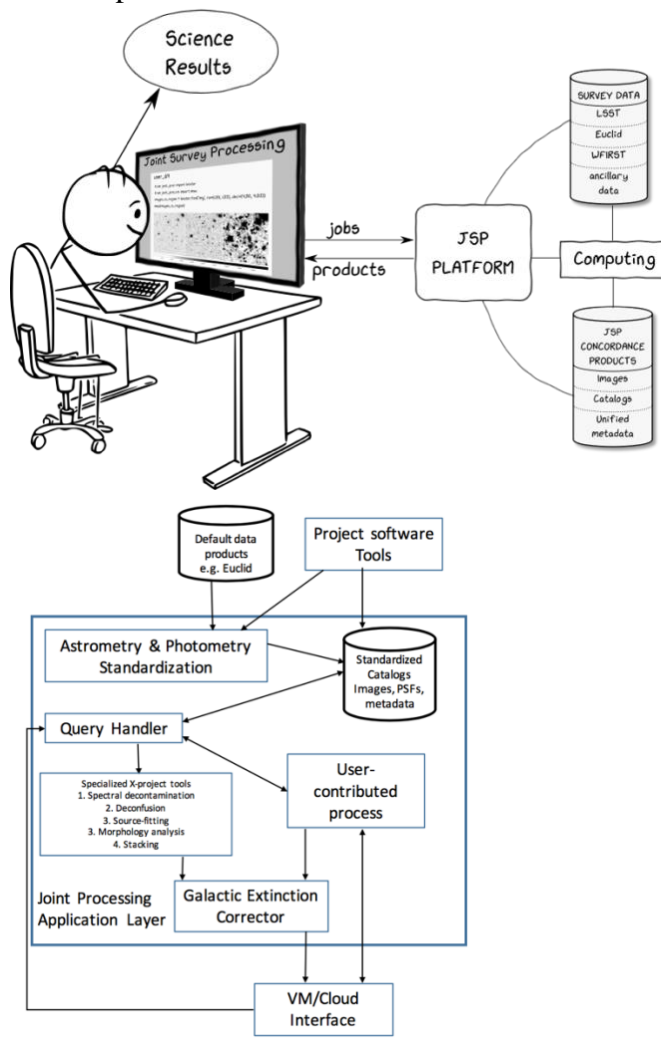

Figure 16: Strawman architecture of the JSP system. The overall layout is shown on the top while the JSP application layer is shown at the bottom. The processing over the entire sky area that overlaps the surveys will be done once and stored in a database for easy retrieval. Agile,

customized processing applications will be executed on-the-fly by code on standardized, calibrated frames from the projects which will be stored at their corresponding data processing centers.

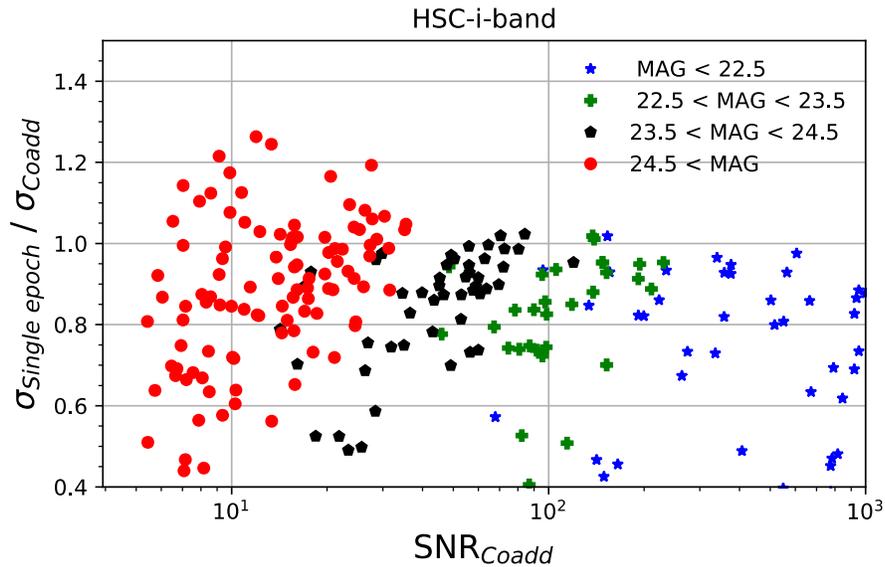

Figure 17: (Top) Improvement in photometric uncertainty of color-selected AGN by using error-weighted single-frame photometry relative to photometric uncertainty derived from a coadded image, based on 0.6" seeing Subaru/HSC data. The improvement in photometric uncertainty is about 15%, ranging from 0.85+/-0.2 at the faint end to about 0.77+/-0.2 at the bright end, suggesting that the photometry derived from single frame LSST data is more accurate than that derived from coadds with the impact slightly more substantial at the bright end.

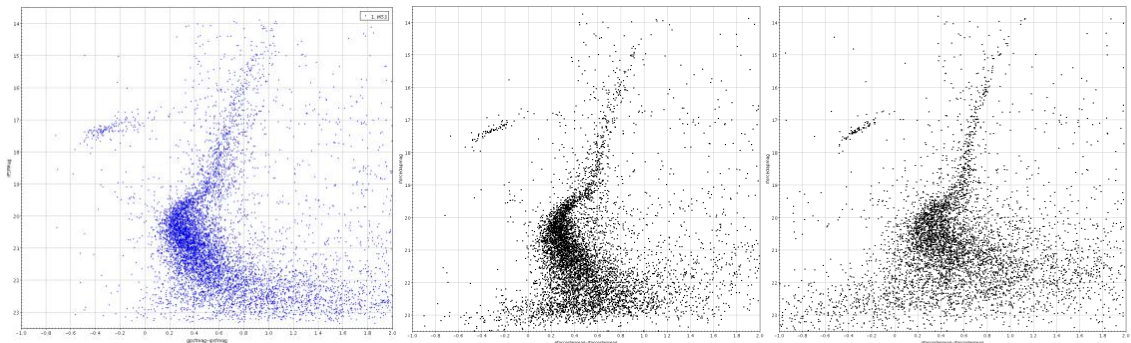

Figure 18: Comparison in derived color-magnitude (g-i vs i) diagrams of stars using different photometric techniques (Metcalfe et al. - Link). This is in a globular cluster (M53) from seeing-limited Pan-STARRS1 data. The left panel shows the result of PSF fitting on a coadded image while the middle panel shows the result of PSF fitting single-epoch images. The right panel shows the result of aperture photometry on a coadded image. Clearly, the middle panel results in the cleanest measure of stellar photometry arguing in favor of PSF fitting analysis on single-epoch images for compact sources.

# 5. Overview of JSP Software Elements

There software development for JSP can be broken down into those elements needed to create the infrastructure, and elements that are related to data manipulation.

The infrastructure software includes 1) setting up and maintaining software containers which can allow data access and visualization, 2) managing network throughput to ensure timely transfer of data from storage sites to high performance compute nodes, 3) extracting unified metadata from the individual datasets, 4) providing seamless over-the-network data access from multiple archives through a data broker, 5) data visualization tools, 6) security and privacy software to ensure that a) a user does not overwhelm computational resources, b) a malicious user does not affect the contents of the archives, c) the work of an individual user is private and cannot be viewed by others and d) secure token based passing of data from archives to containers at external compute nodes.

The data manipulation software includes 1) align and standardize the reduced data products, 2) generate the concordance optical/NIR images/catalogs and 3) allow custom manipulation of these concordance datasets e.g. custom coadds, non-sidereal stacking, point source fitting 4) generate extinction maps and color corrector tool, 5) checking cross-calibration and astrometry, 6) injecting fake sources of different shapes after convolving with the corresponding point spread function. The data broker and client library that we will design will significantly lower the barrier to entry for algorithm developers wishing to co-analyze multi-mission, multi-wavelength data. This data broker would allow users to query by position (or survey object id) on the sky, specify a wavelength range and spatial extent, and have returned to them images (single visit or co-added) aligned to a common grid (likely Gaia) plus metadata drawn from the archives of appropriate surveys (e.g. LSST, WFIRST, GALEX). The metadata supplied with these images would include precise astrometric mapping and a model of the PSF interpretable by JSP algorithms.

A more exhaustive listing of the JSP software that was used in scoping the effort can be found in Table 3 below.

# 6. Cyclic Releases and Quality Control

For every data release from either Euclid, LSST or WFIRST, joint survey processing will be run in bulk mode on the released data products on the overlapping area to generate a new version of the concordance products and supporting software. There will therefore be a mapping between version number of the JSP system and the version number and data release number of each of the other projects, as relevant. For example, JSP1.0 will be the combination of Euclid Q1 (quicklook release 1), processing version 1.0 and LSST DR1, processing version 1.0. If LSST subsequently releases a delta DR1 with processing version 2.0, then the JSP version will be correspondingly updated. This information will be stored in the documentation, catalog and image metadata released as part of JSP. If the infrastructure is built in time for the first releases, we anticipate a short turnaround time of 3 months for the JSP concordance products to be released. We anticipate these cycles to occur approximately annually, and will coordinate with the projects to avoid closely spaced JSP releases driven by the different release calendars of the projects.

Quality control is a major aspect of JSP since the goal is to provide the highest quality deblended photometric catalogs. Basic quality control tests will be done through comparison with the baseline catalogs released by the individual survey projects and through a comparison with known, well studied sources and fields – effectively a reference library of stars and galaxies. For every JSP release, a community-based team will be motivated to analyze the beta-products with the goal of identifying issues with data quality.

# 7. Computing Needs

Three types of computing runs have been identified. The first is required to produce the concordance products and package them for use by the community. The second type encompasses bulk runs managed by large science collaborations aimed at major science investigations such as Weak Lensing or proper motion searches. The third type accounts for the more "localized" science investigations conducted by the community.

The generation of concordance products will need to be run roughly annually to remain up-to-date with releases from the projects. They would start in early 2023 as soon as overlapping data from both LSST and Euclid are publicly available. We estimated in section 4.4 that this effort would involve a total of about 330 million CPU hours, combining both products and supporting simulations. However, each iteration of this for the intermediate data releases is likely only about a tenth of that since the overlapping area will be small. It is reasonable to assume that this could be conducted at one of the DOE supercomputing centers as a contribution to the JSP effort.

Bulk computing runs for major investigations are also likely to be needed roughly annually starting in late 2023 as surveys cover more sky or go deeper. We assume these would be run at Super Computer Centers or potentially on commercial Clouds if the costs are reasonable. These runs would have to be scheduled and staged, with test runs to validate the resourcing leading up to production runs. They would be coordinated with the projects, since they will require streaming major amounts of data out of the project archives, and may impact on-going science operations or frustrate access by other users. The resources required for these computing runs, estimated at tens of millions of CPU hours, will be procured by the investigation teams and are not budgeted here.

The science investigations conducted on a more modest scale by community research groups will run wherever those investigators can accommodate their tailored containerized computing runs. We do not propose a centralized facility to host this computing, and assume the individual investigations will be funded to cover computing costs.

# 8. Tasks and Schedule

Table 3, Tasks and Milestones, shows a draft schedule, summarizing the prime mission span for each of the three projects, then listing each JSP task and its period of performance. The anticipated release dates for data from each of the three projects are shown in those lines, with DR for Data Release and QR for Quick-look Release. Active periods of JSP tasks are color-coded for development, maintenance and operations, and their nominal delivery dates and milestones are indicated by diamonds, and are tied to data releases by the three projects. This draft schedule recognizes the focused efforts required leading up to data releases after the launch of Euclid and start of full survey operations of LSST in 2022, and similarly tied to the launch of WFIRST in 2025. However, no special effort is shown for potential final deliveries by WFIRST of survey data at the end of the prime mission. FY32 as a final year in this estimate is somewhat arbitrary.

The tasks are divided into three categories, with Infrastructure corresponding broadly to Tier 0 above and to other utility functions, Concordance Images and Catalogs corresponding to Tier 1 and Tier 2, and Ancillary Science Tools corresponding to Tier 3.

The first task shown in Table 3, with relatively short duration, is the prototyping effort using data from Hubble/ACS and Subaru/Hyper-SuprimeCam that cover over 2 square degrees in common. These data are fairly accurately representative of data which will flow from WFIRST and LSST respectively. They are therefore ideal testbeds for building up JSP capabilities and improving our understanding of the challenges and opportunities and of the scope of effort required. We have archived the individual calibrated images from these data sets and started building the software infrastructure around them.

## Tasks and Milestones

| Task | Category | FY19 | FY20 | FY21 | FY22 | FY23 | FY24 | FY25 | FY26 | FY27 | FY28 | FY29 | FY30 | FY31 | FY32 |
|---|---|---|---|---|---|---|---|---|---|---|---|---|---|---|---|
| Subaru/HSC-HST/ACS | | | | | | | | | | | | | | | |
| LSST | | | | | | DR1 | | | | | | | | | DR11 |
| Euclid | | | | | | QR1 | DR1 | QR2 | DR2 | QR3 | QR4 | DR3 | | | |
| WFIRST | | | | | | | | | | | | | | | |
| Prototyping with HST/Subaru data | Infrastructure | | | | | | | | | | | | | | |
| Setting up and maintaining Containers | | | | | | | | | | | | | | | |
| Testing Containers on cloud/HPC | | | | | | | | | | | | | | | |
| Database development and maintenance | | | | | | | | | | | | | | | |
| Adapting Visualization Interface | | | | | | | | | | | | | | | |
| Security & privacy , load management | | | | | | | | | | | | | | | |
| High performance networking infrastructure | | | | | | | | | | | | | | | |
| Ingest data and generate indexed metadata | | | | | | | | | | | | | | | |
| Data access software | Concordance Images/Catalogs | | | | | | | | | | | | | | |
| Astrometric offset calibration and correction | | | | | | | | | | | | | | | |
| Photometric consistency and color corrections | | | | | | | | | | | | | | | |
| Extended source fitting and Residual Images | | | | | | | | | | | | | | | |
| Software to generate and ingest concordance catalogs | | | | | | | | | | | | | | | |
| Synthesize and ingest concordance products | | | | | | | | | | | | | | | |
| High resolution Galactic extinction calculator | | | | | | | | | | | | | | | |
| Monte Carlo simulations for catalog validation | | | | | | | | | | | | | | | |
| Interface to external datasets e.g. GALEX, WISE, PS, ZTF | Ancillary Science Tools | | | | | | | | | | | | | | |
| Selective and Non-sidereal image stacking | | | | | | | | | | | | | | | |
| Point source selection and fitting tool | | | | | | | | | | | | | | | |
| Tools for cross-project shape measurements | | | | | | | | | | | | | | | |
| Spectrophotometric deblending tool | | | | | | | | | | | | | | | |

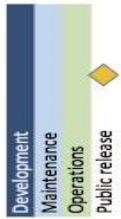

Development
Maintenance
Operations
Public release

Table 3: Task Development Schedule with Delivery Milestones

# 9. Scope Estimate

Table 4 Tasks and Associated Work Effort, shows our estimated scope of effort required to achieve the objectives of the tasks listed above. The tasks listed are the same as in Table 3, shown along with their color-coded phases. An estimate of effort in work-years has been added to each year for each task. The last three columns in this table give a notional distribution of effort across the JSP Core Team, the three survey projects and the research community.

There are two main objectives for JSP, the first being to provide the concordance images and catalogs in the overlapping areas of sky and the second to provide a science platform to the user enabling astronomical manipulation and customized analysis of the concordance products. These goals are reached by implementing tasks in categories Infrastructure and Concordance Images/Catalogs, so it is reasonable to allocate those costs to the JSP and survey projects. The Ancillary Science Tools category is aimed at supporting or developing additional tools for research, and is therefore allocated primarily to the community, with some support by JSP. The items listed in this last category address the most obvious research questions and are provided here for illustration; they are a lower bound on the amount of research activities enabled by JSP.

Summing over the years FY19 to FY24, we estimate a total effort of 100 WY. For FY25-FY32, the total effort is 104 WY. Weighting by the effort distribution in each task, and summing over all years, we estimate a total of 202 WY, 69% of which is allocated to JSP itself, 15% to the individual survey projects, and 16% to the community. The 204 WY number is largely driven by the duration of over a decade. During the five peak years, the effort averages about 24 WY/year for all categories and sources of funding (Table 4 and Figure 19).

A reasonable scenario is that community effort would be funded by proposal and peer-review rounds, with NSF carrying a major part. The individual survey project participation would be funded by their respective funding agencies based on requests by each project. The JSP Core Team could be constituted through direct funding by NASA to one or more of its centers, with contributions from DOE and NSF in the form of dedicated funds or of personnel direction.

A major cost component for JSP is computing, as detailed in section 4.4 above. We estimate a billion or more CPU hours will be needed. A reasonable scenario would be for DOE to allocate that computing load on their Supercomputing Centers, or procure the equivalent on the commercial cloud if that became competitively priced in the next few years. This would motivate more engagement in JSP by DOE laboratory staff, thus helping the overall effort.

In scoping these tasks, we have relied on the experience of the co-authors as leaders and contributors to many science and software development efforts over many years. The primary technique used is scaling from past similar tasks. The prototype work, even though just started, has been very valuable in helping to identify infrastructure components and data manipulation algorithms needed, and their availability. We do not address hardware, procurements or travel expenses here.

## Tasks and Associated Work Effort

| Task | Category | FY19 | FY20 | FY21 | FY22 | FY23 | FY24 | FY25 | FY26 | FY27 | FY28 | FY29 | FY30 | FY31 | FY32 | JSP Project | Survey Projects | Science Comunit |
|---|---|---|---|---|---|---|---|---|---|---|---|---|---|---|---|---|---|---|
| | | | | | | | | | | | | | | | | **Notional Split of Effort** | | |
| Subaru/HSC+HST/ACS | | | | | | DR1 | DR1 | DR2 | DR3 | DR4 | DR3 | | | | 100% | | |
| LSST | | | | | | | | | | | | | | | 100% | | |
| Euclid | | | | | | | | | | | | | | DR11 | 75% | 25% | |
| WFIRST | | | | | | | | | | | | | | | 100% | | |
| Prototyping with HST/Subaru data | Infrastructure | 2 | 1 | 0.5 | 0.2 | | | | | | | | | | | 100% | | |
| Setting up and maintaining Containers | | 1 | 2 | 2 | 2 | 2 | 0.1 | 0.1 | 1 | 0.1 | 0.1 | 1 | 0.1 | 0.1 | 0.1 | 100% | | |
| Testing Containers on cloud/HPC | | | 0.2 | 0.2 | 0.2 | 0.2 | 0.2 | 0.2 | 0.2 | 0.2 | 0.2 | 0.2 | 0.2 | 0.2 | 0.2 | 75% | 25% | |
| Database development and maintenance | | | | 2 | 2 | 0.2 | 0.2 | 0.2 | 0.2 | 0.1 | 0.1 | 0.1 | 0.2 | 0.1 | 0.2 | 100% | | |
| Adapting Visualization Interface | | | 1 | 1 | 0.2 | 0.2 | 0.1 | 0.2 | 0.2 | 0.1 | 0.1 | 0.1 | 0.1 | 0.1 | 0.1 | 100% | | |
| Security & privacy, load management | | | | 0.5 | 2 | 0.6 | 2 | 2 | 0.2 | 0.2 | 0.1 | 0.1 | 0.1 | 0.1 | 0.1 | 100% | | |
| High performance networking infrastructure | | | | 0.2 | 0.4 | 0.6 | 1 | 1 | 1 | 0.5 | 0.5 | 0.5 | 0.5 | 0.5 | 0.5 | 100% | | |
| Ingest data and generate indexed metadata | | | | 1 | 1 | 2 | 3 | 3 | 1 | 3 | 2 | 2 | 2 | 2 | 2 | 50% | 50% | |
| Data access software | Concordance Images/Catalogs | 0.3 | 1 | 2 | 1 | 3 | 1 | 3 | 2 | 2 | 1 | 0.5 | 0.5 | 0.5 | 0.5 | 80% | 20% | |
| Astrometric offset calibration and correction | | 0.3 | 0.5 | 1 | 2 | 3 | 1 | 0.5 | 1 | 0.2 | 0.5 | 0.1 | 0.1 | 0.1 | 0.1 | 75% | 25% | |
| Photometric consistency and color corrections | | 0.2 | 1 | 1 | 3 | 3 | 1.5 | 1.5 | 1 | 1 | 1 | 0.4 | 0.4 | 0.4 | 0.4 | 75% | 25% | |
| Extended source fitting and Residual Images | | | | 1 | 1 | 1 | 0.5 | 0.2 | 0.4 | 0.4 | 0.2 | 0.2 | 0.2 | 0.2 | 0.2 | 20% | 20% | 60% |
| Software to generate and ingest concordance catalogs | | | | 2 | 2 | 3 | 3 | 3 | 2 | 2 | 2 | 0.5 | 0.3 | 0.2 | 0.2 | 80% | 20% | |
| Synthesize and ingest concordance products | | | | 0.5 | 0.5 | 2 | 1.5 | 1 | 2 | 1 | 1 | 1 | 0.5 | 0.5 | 0.5 | 75% | 25% | |
| High resolution Galactic extinction calculator | Ancillary Science Tools | | 0.5 | 1 | 0.3 | 0.2 | 0.2 | 0.2 | 0.3 | 0.3 | 0.2 | 0.1 | 0.1 | 0.1 | 0.1 | 50% | | 50% |
| Monte Carlo simulations for catalog validation | | | | 1 | 0.4 | 2 | 3 | 2 | 2 | 0.5 | 0.5 | 1 | 0.5 | 0.2 | 0.5 | 50% | | 50% |
| Interface to external datasets e.g. GALEX, WISE, PS | | | 0.6 | 1 | 0.2 | 0.2 | 0.2 | 0.6 | 0.6 | 0.2 | 0.2 | 0.2 | 0.1 | 0.2 | 0.1 | 100% | | |
| Selective and Non-sidereal image stacking | | | | | 1 | 1.5 | 0.8 | 0.4 | 0.4 | 0.4 | 0.5 | 1 | 0.3 | 0.2 | 0.2 | 75% | | 25% |
| Point source selection and fitting tool | | | | 0.5 | 0.5 | 0.3 | 0.4 | 1 | 0.4 | 0.3 | 0.5 | 0.4 | 0.3 | 0.2 | 0.2 | 50% | | 50% |
| Tools for cross-project shape measurements | | | | | 2 | 2 | 1 | 0.4 | 1.5 | 1.5 | 0.5 | 0.4 | 0.3 | 0.2 | 0.2 | 20% | 0% | 80% |
| Spectrophotometric deblending tool | | | | | 2 | 2 | 1.5 | 0.5 | 1.5 | 1.5 | 0.5 | 0.4 | 0.3 | 0.2 | 0.2 | 20% | 0% | 80% |
| JSP Project | | 3.62 | 6.93 | 11.95 | 16.13 | 17.93 | 16.00 | 16.22 | 14.70 | 9.83 | 6.87 | 7.93 | 4.50 | 4.08 | 4.08 | 140.74 | | |
| Survey Projects | | 0.19 | 0.58 | 0.90 | 2.68 | 4.15 | 3.80 | 3.49 | 3.75 | 2.93 | 2.07 | 1.72 | 1.45 | 1.43 | 1.43 | | 30.55 | |
| Science Community | | 0.00 | 0.30 | 1.55 | 3.10 | 5.53 | 4.30 | 3.09 | 5.25 | 3.34 | 1.47 | 1.96 | 1.15 | 0.79 | 0.79 | | | 32.62 |
| TOTAL | | 3.80 | 7.80 | 14.40 | 21.90 | 27.60 | 24.10 | 22.80 | 23.70 | 16.10 | 10.40 | 11.60 | 7.10 | 6.30 | 6.30 | **TOTAL=204 WY** | | |

Table 4: Tasks and associated work effort by fiscal year

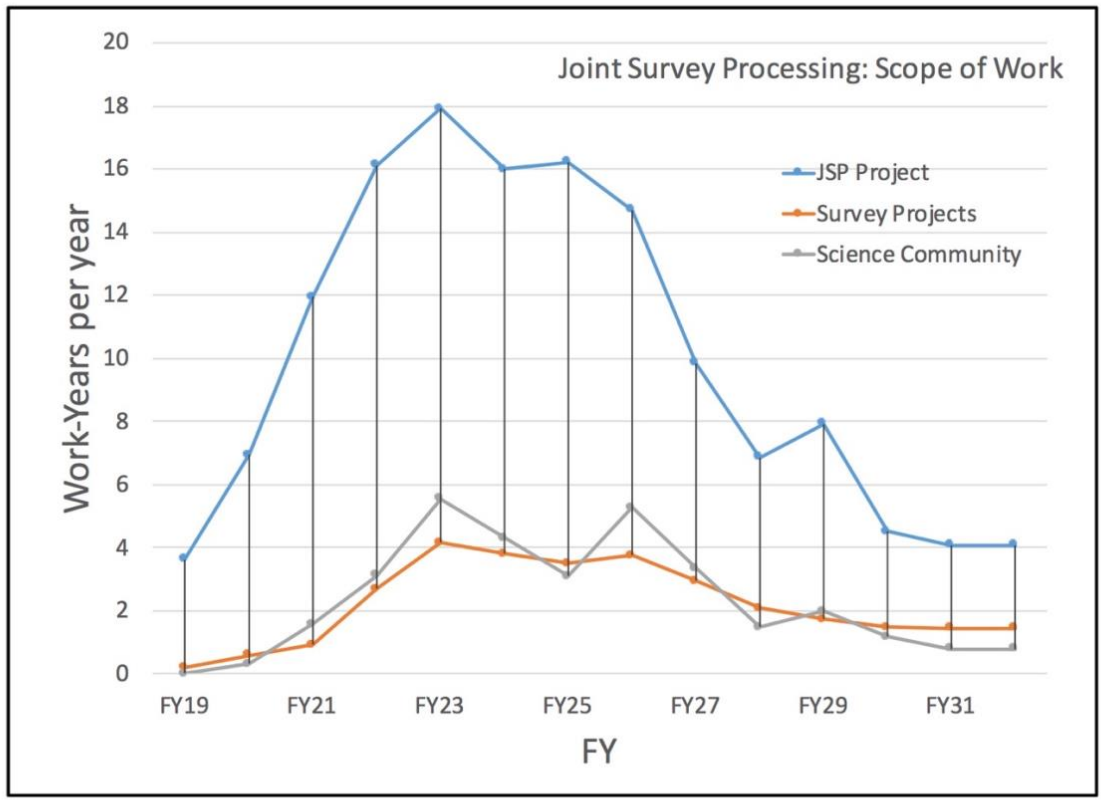

Figure 19. Workforce profile for JSP tasks per team, as shown in Table 4 and discussed in Sec 9.

# 10.   Management and Coordination

The starting point of designing a JSP workflow is to establish what each of the projects (Euclid, LSST and WFIRST) is delivering, on what timescale, the features in each of the datasets, what software exists and what software needs to be developed, augmented, or adapted. Based on the current information available for each of the three projects, the additional effort required at each of the projects to accommodate JSP, while critical, is quite modest compared to their own data processing and management effort, as shown in Table 4.

Timely and efficient information flow among the JSP team and the three projects and their archives as well as archives storing the ancillary datasets will be essential. This will ensure quality and robustness for the deliverables, and help avoid duplication of effort. We propose an eight-member coordination working group (CWG) composed of a scientific and technical representative from each of the three projects and from the JSP team. As needed, the CWG would be augmented with one scientific liaison for the ancillary datasets and one overall archive/technical contact to the ancillary data archives. The main function of this CWG would be to collate and analyze information about status and schedule on data products, services and infrastructure feeding into JSP, to enable the most efficient use of resources in meeting JSP deliverables.  The CWG reports the results of their analysis against the baseline JSP schedule, and recommends actions (if issues arise) to all stake holders (projects, ancillary data archives and JSP team) at least once a quarter.

Each of the funding agencies (DOE, NASA, NSF) has standard processes for funding and reporting for each of the projects. Once the JSP effort is selected for implementation, these processes would similarly apply to it, in addition to the agencies augmenting the charge and funding to the projects for their participation in the effort. Most of the JSP effort will be conducted by a JSP Core Team (CT) whose management is attached to an institution linked to the funding sources and mechanisms. In order to provide transparency to all three agencies, a JSP management board consisting of the managers or their designees of the JSP-CT and the data systems of each of the projects will convene on a monthly basis to review status and decide on any modifications to the baseline plan, and will jointly report on progress to the agencies on a quarterly basis. The management board meetings will be chaired by the JSP-CT lead.

In addition, we expect to have a Science Steering Committee chosen from the U.S. user community who will identify priorities and assess the value-added scientific capabilities of JSP on an annual basis. A User Committee would review the implementation aspects, especially as they relate to ease-of-use and applicability to pressing issues faced by the research community.

## 12.  Acknowledgements

We are very grateful to Daniela Calzetti, George Djorgovski and Bahram Mobasher for their critical review and suggestions which resulted in a significant improvement of this report.

# 13. Working Group Membership

The table below shows the difference topics that were scoped as part of this document, the leads, and the membership within the group.

| Task | a.i | a.ii | a.iii | a.iv | b.i | b.ii | b.iii | c.i | c.ii | c.iii | c.iv | c.v |
|---|---|---|---|---|---|---|---|---|---|---|---|---|
| | SCIENCE | | | | ALGORITHMS | | | ARCHITECTURE | | | | |
| Topics | Cosmology | Agile Extragalactic Science | Galactic, Solar System, Stellar | Microlensing, time baselines for SSOs | LSST Deconfusion, optimal photometry for phot-z, cross-mission color selection, dust corrections | | | Computing infrastructure, networking, hardware, archives | | | | |
| Lead(s) | J. Newman (UPitt.) | R. Chary (IPAC) | R. Paladini & S. Wachter (IPAC) | G. Helou (IPAC) | H. Ferguson (STScI) | H. Ferguson (STScI) | P. Melchior (Princeton) | | | B. Rusholme (IPAC) & A. Smith (STScI) | P. Appleton (IPAC) | H. Teplitz (IPAC) & A. Smith (STScI) |
| Members | Momcheva, Ferguson, Schneider, Prakash, Chary, Capak | Ferguson, Momcheva, Prakash, Capak, Armus, Wood-vasey, Malhotra, McEnery | Ferguson, Momcheva, Kirkpatrick, Chary, Grillmair | van der Marel, Carey, Grillmair | Dawson, Melchior, Schneider, Schulz B, Lee, Appleton | Dawson, Melchior, Ferguson, Schneider, B. Lee, Grillmair, Armus | Dawson, Ferguson, Koekemoer, Lupton, Schulz | Schneider, Fox, Groom, Ebert | Fox, Flynn, Ebert | Fox, Groom, Berriman | Smith, Fou, Wachter, B. Lee, Rusholme, Berriman | Fox, Wachter, Groom, Rusholme, Berriman |

Table 5: Working Group Membership